\documentclass[11pt]{amsart}
\usepackage{latexsym,amssymb,amsmath,
graphics,
}

\newcommand{\arr}{\longrightarrow}

\renewcommand{\hat}{\widehat}

\newcommand{\mx}[1]{\mbox{{#1}}}
\newcommand{\s}{\quad}

\newcommand{\Sym}{{\tt S}}

\newcommand{\cycl}[3]{\mx{${\sf C}(#1_\bullet #2_\bullet #3_\bullet)$}}
\def\wtg{{\mathrm{wt}_1\,}}
\def\wt{{\mathrm{wt}_2\,}}
\def\wtt{{\mathrm{wt}_3\,}}
\DeclareMathOperator{\sdeg}{\deg_{\Sym}}
\DeclareMathOperator{\ldeg}{\deg_{\La}}
\DeclareMathOperator{\udeg}{\deg_{U}}
\newcommand{\lxl}[1]{{\bf lex}(<#1)}
\newcommand{\lxq}[1]{{\bf lex}(\leq#1)}
\newcommand{\dlxl}[1]{{\bf dlex}(<#1)}
\newcommand{\dlxq}[1]{{\bf dlex}(\leq#1)}

\newcommand{\lsp}[1]{\langle {#1} \rangle}

\newcommand{\dst}{\mbox{{\sf d}}}
\newcommand{\dvar}[1]{\mbox{{\sf d}$x_{#1}$}}
\newcommand{\ddind}[2]{\mbox{{\sf d}$_{#1#2}$}}

\newcommand{\pt}{\mbox{$\partial$}}
\newcommand{\dprt}{\mbox{${\partial}$}}
\newcommand{\hpt}{\mbox{$\hat{\partial}$}}
\newcommand{\dpind}[1]{\mbox{${\partial}_{#1}$}}
\newcommand{\dhind}[1]{\mbox{$\hat{\partial}_{#1}$}}

\newcommand{\Dep}{\mx{$\Delta^{\!\!{}^{\scriptscriptstyle{+}}}\!$}}
\newcommand{\Dem}{\mx{$\Delta^{\!\!{}^{\scriptscriptstyle{-}}}\!$}}
\newcommand{\lag}{{\tt g}}
\newcommand{\gind}[1]{{\tt g}_{#1}}

\newcommand{\al}{\alpha}
\newcommand{\be}{\beta}

\newcommand{\De}{\Delta}
\newcommand{\ep}{\varepsilon}

\newcommand{\ka}{\varkappa}

\newcommand{\la}{\lambda}
\newcommand{\La}{\Lambda}

\newcommand{\si}{\sigma}

\newcommand{\ph}{\varphi}

\newcommand{\tht}{\theta}

\newcommand{\Om}{\Omega}

\def\Bw{{\mathcal B}}

\def\Hw{{\mathcal H}}

\def\Nw{{\mathcal N}}

\def\Sw{{\mathcal S}}

\def\CC{{\mathbb{C}}}

\def\ZZ{{\mathbb{Z}}}

\newcommand{\fg}{\mbox{{\tt g}}}

\renewcommand{\theequation}%
  {\arabic{section}.\arabic{equation}}
\makeatletter
\renewcommand\section%
 {\@startsection {section}{1}{\z@}%
  {-3.5ex \@plus -1ex \@minus -.2ex}%
    {2.3ex \@plus.2ex}%
       {\normalfont\large\bfseries}}
\@addtoreset{equation}{section}
\makeatother

\makeatletter
\renewcommand\subsection%
{\@startsection{subsection}%
{2}%
{1em}
{-\baselineskip}%
{-1ex\@plus -1ex \@minus -.2ex }%
{\normalfont\large\bfseries}%
{}%
}
\makeatother

\newtheorem{Proposition}{Proposition}[section]

\newcommand{\bPr}{\begin{Proposition}}
\newcommand{\ePr}{\end{Proposition}}

\newtheorem{Theorem}[Proposition]{Theorem}
\newcommand{\bTh}{\begin{Theorem}}
\newcommand{\eTh}{\end{Theorem}}

\newtheorem{Lemma}[Proposition]{Lemma}
\newcommand{\bLe}{\begin{Lemma}}
\newcommand{\eLe}{\end{Lemma}}

\newtheorem{Definition}[Proposition]{Definition}
\newcommand{\bDe}{\begin{Definition}}
\newcommand{\eDe}{\end{Definition}}

\newtheorem{Corollary}[Proposition]{Corollary}
\newcommand{\bCo}{\begin{Corollary}}
\newcommand{\eCo}{\end{Corollary}}

\newtheorem{Conjecture}[Proposition]{Conjecture}
\newcommand{\bCj}{\begin{Conjecture}}
\newcommand{\eCj}{\end{Conjecture}}

\newtheorem{Remark}[Proposition]{Remark}

\newcommand{\bEq}{\begin{equation}}
\newcommand{\eEq}{\end{equation}}

\newcommand{\bEa}{\begin{eqnarray}}
\newcommand{\eEa}{\end{eqnarray}}

\newcommand{\bEaz}{\begin{eqnarray*}}
\newcommand{\eEaz}{\end{eqnarray*}}

\newcommand{\bAr}{\begin{array}}
\newcommand{\eAr}{\end{array}}

\newcommand{\bN}{\begin{enumerate}}
\newcommand{\eN}{\end{enumerate}}

\newcommand{\bD}{\begin{description}}
\newcommand{\eD}{\end{description}}

\newcommand{\epf}{$\Box$}

\newcommand{\alphaparenlist}{%
  \renewcommand{\theenumi}{\alph{enumi}}%
  \renewcommand{\labelenumi}{(\theenumi)}%
}

\newcommand{\arabicparenlist}{%
  \renewcommand{\theenumi}{\arabic{enumi}}%
  \renewcommand{\labelenumi}{(\theenumi)}%
}
\newcommand{\romanparenlist}{%
  \renewcommand{\theenumi}{\roman{enumi}}%
  \renewcommand{\labelenumi}{(\theenumi)}%
}

\begin{document}

\title[Representations of $E(3,6)$III: Singular vectors.]
{Representations of the exceptional
Lie superalgebra
$E(3,6)$ III:~Classification of singular vectors.
}

\author{Victor G. Kac
${}^*$
and Alexei Rudakov
}
\thanks{${}^*$~Supported in part by NSF grant DMS-0201017. \\
}


\begin{abstract}
We continue the study of irreducible representations
of the exceptional Lie superalgebra $E(3,6)$. This is
one of the two simple infinite-dimensional Lie superalgebras of vector
fields which have a Lie algebra
$s\ell(3)\times s\ell(2)\times g\ell(1)$ as the zero degree component
of its consistent $\ZZ$-grading.
We provide the classification of the singular vectors
in the degenerate Verma modules over $E(3,6)$, completing thereby
the classification and construction of all irreducible $E(3,6)$-
modules that are $L_0$-locally finite.
\end{abstract}

\maketitle

\section{Introduction.}\label{sec:1}

There are only two  infinite-dimensional
simple linearly compact Lie superalgebras,
$E(3,6)$ and $E(3,8)$, that have the Lie algebra
$s\ell(3)\times s\ell(2)\times g\ell(1)$ as the zero degree component
$\gind 0$ in their consistent $\ZZ$-grading [2]. In the present paper
we continue the study of irreducible representations of $E(3,6)$,
the Lie superalgebra which has apparent relations to the
Standard Model (see [4]).

The article is a sequel to our  papers [3, 4] and together they provide
the classification and description of all irreducible  $L_0$-locally finite
representations of $E(3,6)$ (see [3] for the definition).

As was shown in [3] the problem can be solved in two steps. First
one obtains
the classification of the so called degenerate (generalized) Verma modules
(the Verma modules that have non-trivial singular vectors),
and then one describes the irreducible factors of
these degenerate Verma modules.

In [4], both probelms were solved
modulo the classification of singular vectors.
In this article we provide the proof of this classification (the
result was announced in [4]). Thus we complete
the classification
of the irreducible $E(3,6)$-modules that are $L_0$-locally finite.

As before all vector spaces, linear maps and tensor products  are
considered over the field $\CC$ of complex numbers.

\section{Notations and basic basic properties of $E(3,6)$.}

We recall here the basic definitions and notations from [1,2,3].

To construct $E(3,6)$ we use its embedding into $E(5,10)$.
Namely consider even variables $x_1, \ldots, x_5$. We also use
the notations $z_{+}=x_4$,  $z_{-}=x_5$ further on.
Let $S_5$ be the Lie algebra of  divergence zero formal
vector fields in these
variables, and $\dst\Om^1(5)$ the space of closed (=exact)
formal  differential
$2$-forms in these variables.

Let us remind that the Lie superalgebra $E(5,10)$ has
$E(5,10)_{\bar{0}}\simeq S_5$ as a Lie algebra,
$E(5,10)_{\bar{1}}\simeq\dst\Om^1(5)$
as an $S_5$-module and the brackets on $E(5,10)_{\bar{1}}$ are
defined via the exterior product
of  differential forms.

We shall be using the following notations:
\begin{displaymath}
\ddind jk  := \dvar{j} \wedge \dvar{k}, \quad
\dst^+_i := \dst_{i4} \quad \dst^-_i := \dst_{i5} \qquad
\dpind i  := \dprt /\dprt x_i  \quad
\dpind{+}   :=\dpind 4,\quad \dpind{-} :=\dpind 5 .
\end{displaymath}

An element $A$ from $E(5,10)_{\bar{0}}=S_5$ can be written as
\[ A=\sum_i a_i \dpind{i}\, , \s \mx{ where } a_i \in \CC \left[\left[
  x_1,\ldots,x_5\right] \right] ,\s
 \sum_i \dpind{i}a_i =0 \, ,
\]
and an element $B$ from $E(5,10)_{\bar{1}}$ is of the form
\[ B=\sum_{j,k} b_{jk}\ddind jk \, , \hbox{ where }
b_{jk} \in \CC [[ x_1 , \ldots , x_5 ]], \, dB=0 \, .
\]
The brackets in $E(5,10)_{\bar{1}}$
can be computed using bilinearity and the rule
\[ [a\ddind jk ,b\ddind lm ]= \ep_{ijklm} ab \dpind{i}
\]
where
$\ep_{ijklm}$ is the sign
of the permutation $(ijklm)$ when
$\{i,j,k,l,m\}$
are distinct and zero otherwise. \\

We have for $L = E(3,6)$ the following description
of the first three pieces of its consistent $\ZZ$-grading
$L =\Pi_{j \geq -2}\, \fg_j$:
\begin{displaymath}
\gind{-2}= \lsp{\dpind{i},\, i=1,2,3}, \quad
\gind{-1} =  \lsp{\ddind ij ,\,i=1,2,3,\,j=4,5}.
\end{displaymath}
And
$\fg_0 = s \ell (3) \oplus s \ell(2) \oplus g\ell (1)$
with the following basis:
\begin{eqnarray*}
&&h_1=x_1\dpind 1-x_2\dpind 2, \quad
h_2=x_2\dpind 2-x_3\dpind 3, \quad
e_1=x_1\partial_2, \quad
e_{12}=x_2\partial_3, \quad e_3=x_1\partial_3,\\
&&f_1=x_2\partial_1, \quad \!\!\!
f_2=x_3\partial_2, \quad \!\!\!
f_{12}=x_3\partial_1, \quad \!\!\!
h_3=x_4\dpind 4-x_5\dpind 5, \quad \!\!\!
e_3=x_4\dpind 5,  \quad f_3 =x_5\dpind4,\\
&&Y=\tfrac{2}{3}(x_1\dpind 1+x_2\dpind 2+x_3\dpind 3)-(x_4\dpind 4+x_5\dpind 5).
\end{eqnarray*}

We keep the
standard Cartan subalgebra $\Hw=\lsp{h_1,h_2,h_3,Y}$ and
the standard Borel
subalgebra $\Bw=\Hw \oplus  \Nw $, where
$\Nw =\langle e_1, e_2, e_{12},\,e_3  \rangle$, of $\gind 0$.  We
denote by $\wt v$ the $s\ell(2)$-weight of $v$ and by
$\wtt v$ the $s\ell(3)$-weight whenever the weight is defined.
We denote by $\wtg v$ the eigenvalue of $Y$ on $v$ whenever defined.\\

The algebra $ E(3,6)$ is generated by
$\gind{-1}, \gind 0, \gind 1$
moreover it is generated
by $\gind 0$ and the three elements $e_0, f_0$ and
$e^{-}_0$ (the notation of the last element in [3] was $e'_0$)
from the following four:
\begin{eqnarray*}
f_0 &=&\ddind 14,\\
e^{-}_0 &=& x_3\ddind 35,  \\
e^{+}_0 &=& x_3\ddind 34,  \\
e_0 &=& x_3\ddind 25-x_2\ddind 35+2 x_5\ddind 23 \, .
\end{eqnarray*}
It is known ([3]) that
the element $f_0$ is the highest weight vector of the
$\fg_0$-module $\gind{-1}$, while $e^{-}_0, e_0$ are the lowest
weight vectors of the $\fg_0$-module $\gind{1}$  and
 \begin{eqnarray}
   \label{eq:1.1}
   [e^{-}_0,f_0] &=& f_2, \\
   \label{eq:1.2}
   [e_0,f_0] &=& \tfrac{2}{3}h_1+\tfrac{1}{3}h_2- h_3 - Y =: h_0.
 \end{eqnarray}

We use the notations
$\lag^{\pm}_{-1}=\lsp{\dst^{\pm}_1,\dst^{\pm}_2,\dst^{\pm}_3}$,
and the following ones for the elements
from the exterior algebras
$\La^{\pm}:= \La\lag^{\pm}_{-1}$:
\begin{displaymath}
\dst^{\pm}_{ij} := \dst^{\pm}_i \cdot \dst^{\pm}_j, \qquad
\dst^{\pm}_{ijk} := \dst^{\pm}_i \cdot \dst^{\pm}_j\cdot \dst^{\pm}_k \, .
\end{displaymath}

It is easy to see ([3]) that
the following subalgebras  $\fg^{\pm}_1$ are abelian and are
normalized
by $s\ell (3)$:
\begin{equation}\label{eq:1.3}
  \fg^\pm_1 = \langle x_i\dst_{j}^\pm + x_j \dst_{i}^\pm |\, i,j=1,2,3 \rangle \,
  .
\end{equation}
Note that $ [\fg^{\pm}_{-1}, \fg^{\mp}_1] = 0 $,
and that the subalgebras
\begin{displaymath}
  \Sw^{\pm} = \fg^{\pm}_{-1} \oplus s\ell (3) \oplus \fg^{\pm}_1 \, .
\end{displaymath}
are each isomorphic to the simple Lie superalgebra $S(0|3)$ of
divergenceless vector fields in three anticommuting
indeterminates.

It is important for us that
the change of the variables $x_i \to x_i, \,\,i\leq 3$, $x_4 \to x_5$,
$x_5 \to x_4$ induces an automorphism $\ph$ of the algebra
 $ L=E(3,6)$.
Clearly  $\ph$ preserves the grading and the subalgebras $s\ell (3)$,
$s\ell (2)$, $g\ell (1)$
of $\fg_0$. The restriction of $\ph$ on $s\ell (3)\oplus g\ell (1)$
is the identity
map, but on $s\ell (2)$ we have
\begin{equation}\label{eq:1.4'}
\ph \,e_3=f_3, \s \ph \,f_3=e_3, \s\ph \, h_3=-h_3\,.
\end{equation}
Also $\ph$ interchanges $\Sw^{\pm}$ and
its action  on $L_-$ is defined by the formulae
\begin{equation}\label{eq:1.4}
\ph \, \dst^{\pm}_i =  \dst^{\mp}_i\, ,\s \ph \,\dhind i = - \dhind i\,
\s \text{ for } i=1,2,3\,.
\end{equation}

\section{Contraction and quasi-singular vectors.}
\label{sec:2}

We use notation
$\,\,L_-=\oplus_{j<0} \,\fg_j,\quad L_+=\Pi_{j>0}\, \fg_j,\quad
L_0 =\fg_0 \oplus L_+\,.\quad$
Of course
  \begin{displaymath}
    L=L_- \oplus \fg_0 \oplus L_+ \, ,
  \end{displaymath}
Our main objective is to study singular vectors in the generalized Verma modules
\begin{equation}
\label{eq:2.1}
  M(\mx{\bf V}) = U(L ) \otimes_{U(L_0)}\!\mx{\bf V} \cong
U (L_-) \otimes_{U(\fg_0)}\!\mx{\bf V}  \, ,
\end{equation}
where as usual $\mx{\bf V}$ is a finite-dimensional irreducible $\fg_0$-module
extended to $L_0$ by letting $\fg_j$ for $j>0$ acting
trivially.\\

We are concerned with singular vectors in $M(\mx{\bf V})$
that are also the highest weight vectors with respect to the
standard Cartan and Borel subalgebras $\Hw$ and  $\Bw$ of $\gind 0$.
By Proposition 2.2 of [3]
the defining property of these vectors can be written as follows:
\begin{equation}
\label{eq:2.2}
\begin{array}{l}
\text{\it The $\fg_0$-highest weight vector  $\mx{ \bf v}$ of a
\mx{$E(3,6)$}-module
is  {\sl singular} iff}\\
 \qquad\quad e_0 \cdot \mx{\bf v} =0, \qquad e^{-}_0 \cdot \mx{\bf v} =0 \, .
\end{array}
\end{equation}
%
Property  $e^{-}_0 \cdot \mx{\bf v} =0$
implies $e^{+}_0 \cdot \mx{\bf v} =0$ when
$\mx{\bf v}$ is an $s\ell(2)$-highest weight vector,
but in general these are different
conditions. For $s\ell(3)$-highest weight vectors
the first amounts to being $\Sw^-$-singular (i.e. killed by
$ \fg^{-}_1 $)
and the second to being $\Sw^+$-singular
(i.e. killed by
$ \fg^{+}_1 $).

%
\begin{Definition}\label{D:2.1}
We call  a vector $v$ {\sl quasi-singular vector} if it
is $s\ell(3)$-highest weight vector
and $\Sw^-$-singular.
We call a vector $v$ {\sl semi-singular vector}
if it is $s\ell(3)$-highest weight vector
and  is both $\Sw^-$ and $\Sw^+$-singular.
\end{Definition}
In the sequel we shall often say {\it highest vector} in place of
{\it highest weight vector}.\\
Of course an $s\ell(3)$-highest singular vector is both semi-singular
and quasi-singular. Also the $\fg_0$-highest vector of an
$E(3,6)$-module is singular if only only if it is quasi-singular and
is annihilated by $e_0$. But for a vector $\mx{\bf v}$
that is not $\fg_0$-highest
the property  $e^{-}_0 \cdot \mx{\bf v} =0$ does not in general
imply $e^{+}_0 \cdot \mx{\bf v} =0$, the condition of being
quasi-singular is indeed weaker.

We will first describe quasi-singular vectors,
then narrow the ``list of suspects'' to semi-singular ones,
and then come to the description of the $\fg_0$-highest singular vectors.

It is important to mention that we can consider quasi-singular vectors
in a space that is not necessary $E(3,6)$-module, but only
$\Sw^-$-module.\\

For a vector space $T$ and
 a linear form $\tau:T\arr \CC$, $\,\,\tau(t)=
\langle t|\tau\rangle$,
we shall use the contraction maps defined as follows.
For any
linear space $A$  the
contraction map $\langle \tau\rangle: A\otimes T \arr A$
applied to  $a\otimes t$ gives $\langle t|\tau\rangle a$.
\begin{Remark}
\label{R:2.2}
{\sl
If  both $A$ and $T$ are $s\ell(2)$-modules,
$w \in A\otimes T$ is a weight vector of weight $\la$ and  a form $\tau$
has weight $\mu$, then the contraction $w\langle \tau\rangle$ has
weight $\la + \mu$.
When $w$ and $\tau$  are eigenvectors for $Y$, the contraction $v=w\langle \tau\rangle$ is
also an eigenvector (with eigenvalue equals to the sum of the eigenvalues).
}
\end{Remark}

Let $\mx{\bf V}$ be a (finite-dimesional)  $\fg_0$-module
that is isomorphic
to the tensor product of  $s\ell(3)\oplus g\ell(1)$-module $V$
and  $s\ell(2)$-module $T$ (i.e.$\mx{\bf  V}=V\otimes T$ ).
Denote by $F(p,q)$ the irreducible
representation of
$s\ell(3)$ with the highest weight $(p,q)$.
\begin{Remark}
\label{R:p-or-q}
{\sl
The main theorem from [3] states that the non-trivial highest singular vector
$\mx{\bf v} \in M(\mx{\bf V})$ could exist
only when the $s\ell(3)$-module $V$
is a submodule  of $\CC[x_1,x_2,x_3 ]$ or $\CC[\,\dpind1, \dpind2, \dpind3]$.
Therefore we shall look for  quasi-singular and semi-singular vectors
only in $U(L_-)\otimes_\CC \mx{\bf V}
$,  where
$\mx{\bf  V}=V\otimes T$, and ${V}$
belongs to
$\CC[x_1,x_2,x_3 ]$ or $\CC[\,\dpind1, \dpind2, \dpind3]$, or for that
matter we can suppose that either $V=F(p,0)$ or $V=F(0,q)$.
}
\end{Remark}

Let us notice that there is a natural $g\ell(1)$-action on
$\CC[x_1,x_2,x_3 ]$ and $\CC[\,\dpind1, \dpind2, \dpind3]$
such that
\begin{equation}
  \label{eq:Yact}
  Y\,x_i=\tfrac{2}{3}x_i\,, \s Y\,\pt_i=-\tfrac{2}{3}\pt_i\,.
\end{equation}

\begin{Remark}
\label{R:gweit}
{\sl
We will assume that $s\ell(2)$
acts  trivially
on  $V$ and
$g\ell(1)$ acts according to the above rule,
thus that
$\wt$, $\wtg$ are defined on the tensor product $M(V)=U(L_-)\otimes V$.
}
\end{Remark}

Let us notice that the tensor product of $s\ell(3)$-modules $U(L_-)\otimes V$
has natural \mbox{$\Sw^\pm$-structures} and that there is a natural
isomorphism
\begin{equation}
  \label{eq:VTiso}
  M(\mx{\bf V})\cong (U(L_-)\otimes V) \otimes T
\end{equation}
so {\it the contraction map } for any $\tau \in T^*$ is well defined.
We shall use this map further on.
It is of course the homomorphism of $\Sw^-$ (or $\Sw^+$)-modules.
The following result will play a key role in our subsequent
calculation of singular vectors.
\begin{Proposition}\label{P:2.3}
Given
a non-zero $\fg_0$-highest singular vector
$\mx{\bf v}\in M(V\otimes T)$ consider a contraction
$v=\mx{\bf v}\langle \tau\rangle\in U(L_-)\otimes V$, where $\tau \in T^*$.
\begin{enumerate}
\item For any $\tau$,  the contraction $v$ is a semi-singular vector.
\item Whenever $\tau$ is an $s\ell(2)$-weight vector, $v$ is
such a vector too.
\item There exists $\tau \in T^*$
 such that
$v$
is a non-zero semi-singular vector of non-negative \mx{$s\ell(2)$-weight},
and $v$ is also a $g\ell(1)$-weight vector.
\end{enumerate}
\end{Proposition}
\begin{proof}
Whatever $\tau$, the contraction clearly commutes with the
$s\ell(3)$-action, therefore we get the $s\ell(3)$-highest vector
from $s\ell(3)$-highest one.

Because
\begin{equation}
  \label{eq:2.3}
 [e^{\pm}_0,\,\fg_{-1}] \subset s\ell(3) \subset \fg_0
\end{equation}
the $\Sw^+$ and $\Sw^-$-structures are defined on
$U(L_-)\otimes V$.
The operators $e^{\pm}_0$ act on both sides of the contraction map,
and commute with the map . Thus $v$
will be annihilated by $e^{\pm}_0$ because $\mx{\bf v}$ is, hence $v$ is
semi-singular.

To finish the proof, notice that
for any  non-zero $s\ell(2)$-highest vector in $A\otimes T$
there always
exists a contraction such that the result is a non-zero
vector of non-negative weight. This follows immediately
from the formula for the highest weight vectors in the
tensor product of irreducible (finite-dimensional)
$s\ell(2)$-modules.
\end{proof}

\begin{Definition} We say that a semi-singular
(resp. quasi-singular) vector is admissible if the
vector has a non-negative $s\ell(2)$-weight.
\end{Definition}
Proposition~\ref{P:2.3} and Remark~\ref{R:p-or-q} show
that we should
look for admissible quasi-singular and semi-singular vectors
 in $U(L_-)\otimes_\CC V$ where $V$ belongs to
$\CC[x_1,x_2,x_3 ]$ or $\CC[\,\dpind1, \dpind2, \dpind3]$.
We provide the description of these vectors in the next section
after introducing appropriate notations.
\begin{Remark}\label{R:ass}
{\sl
Let us notice that
the authomorphism $\ph$ naturally extends to $U(L_-)\otimes_\CC V$, where
it acts on $U(L_-)$ according to formulae \eqref{eq:1.4} and on $V$
identically. Evidently a vector $\ph(w) \in U(L_-)\otimes_\CC V$
is  semi-singular, if and only if
$w$ is semi-singular. Because of \eqref{eq:1.4'} we get
$\wt w = - \wt\ph(w)$.
Therefore applying $\ph$ we shall get
all semi-singular vectors as soon as we know the admissible ones.
}
\end{Remark}

\section{The description of admissible quasi-singular and
semi-singular vectors.}
\label{sec:3}

We need elaborated notations and some lemmas
before starting the explicit calculation of the quasi-singular
and semi-singular vectors.
We keep the following notations:
\[
 \Sym^k:=\text{Sym}^k(\gind{-2})\, ,\quad
 \Sym=\sum_{k \geq 0} \Sym^k, \quad
\La^{\pm}_i:= \La^i(\gind{-1}^{\pm}).
\]
As in [3] we use ``a hat'' to mark the elements of
$\fg_{-2}\subset \Sym \subset U(L )$, thus further on
\begin{equation*}
\Sym=\CC[\,\hat{\partial}_1, \hat{\partial}_2, \hat{\partial}_3].
\end{equation*}
We will follow the approach developed in [3] for working with
the highest  vectors in tensor products of $sl(3)$-modules where
the first factor is $\Sym$. \\

Let us
refine the isomorphism in \eqref{eq:2.1}
to
an $s\ell(3)$-isomorphism of $M(V)$ with the tensor products of $s\ell(3)$-modules
(we omit the tensor product signs):
\begin{eqnarray}
\label{eq:b-is}
M(V)\cong \Sym\, \La^-\La^+\, V.
\end{eqnarray}
We use this isomorphism later on.
Also for  later use we introduce the following notations.

We denote $\ldeg v$ the combined degree of $\La^-\La^+$-part
in $w\in \Sym\, \La^-\La^+\, V$, for example
$\ldeg \dhind{}_1\dhind{}_2\, \dst_1^-\dst_{12}^+\,v =3$.
Let us mention that $\wt w$ depends only on the $\La^-\La^+$-part of $w$,
for example
$\s
\wt \dst_1^-\dst_{12}^+\,v=
\wt \dhind{}_1\dhind{}_2 \,\dst_1^-\dst_{12}^+\,v =
+1$.

For any space $M$ given with the isomorphism $M\cong \Sym\otimes W$,
whatever $W$, we can think of an element $f\in M$ as a polynomial in
$\hat{\partial_i}$ with coefficients from $W$ written to the right of these
``variables''. We define  $\sdeg f$ to be the {\it degree} of such a
polynomial $f$
.
We also define subspaces $\,\,\lxl{a}M$ (resp. $\lxq{a}M$),
for a multi-index $a$,
to be the linear span of monomials $\,\hat{\partial}^b\cdot w$,
any   $w\in W$ and $\,\,b<_{\mx{\small{lex}}}a\,$,
(resp. $b\leq_{\mx{\small{lex}}}a$).
As usual the expression
\[
  f \equiv g \quad\bmod \lxl{a}
\]
means that $f-g\in  \lxl{a}M$, and we omit $M$ when it is clear from the
context. Similarly
\[
  f \equiv g \quad\bmod \sdeg (<N)
\]
means that $\sdeg(f-g)<N$.

As in [3] we say that
$\,\hat{\partial}^a\cdot w\neq 0$ is
the {\it lexicographically highest term } of $f$ iff
\[ f \equiv \hat{\partial}^a\cdot w \quad\bmod \lxl{a}.\]
Similar notations will be used for the {\it degree-lexicographic}
order on the monomials that is
defined by the condition
\begin{equation}
  \label{eq:3.1}
  \partial^b\leq_{\mx{\small{dlex}}}\partial^a \iff (|b|,b)
\leq_{\mx{\small{lex}}}(|a|,a),
\quad\text{ where $|a|=\sdeg\partial^a=\sum a_i$}.
\end{equation}


Let $s$ be an integer, denote
$h'\{s\}=h_1+h_2+s+1$, $ h_2\{s\} = h_2+ s$.
Slightly generalizing the definition in Section 3 of [3] let
 $B,\, D_i\{s\}\in \Sym \# U( sl(3))$ be:
  \begin{align}                                       
 B\,\,\,\, &= f_{12}h_1+f_2f_1= B = f_{12}(h_1+1)+f_1f_2,\notag\\
D_1\{s\}&=\dhind1\,h_1h'\{s\} +
\dhind2\, f_{1}h'\{s\}  +
\dhind3\,B
,\notag
\\
D_2\{s\}&=\dhind2\,h_2\{s\} +\dhind3\, f_2,
\label{eq:D}\\
D_3\{s\}&=\dhind3 \notag.
\end{align}
We write simply $D_i$ when $s=0$. We also use the usual multi-index
notations
\[
D^a\{s\}:=D\{s\}^a=D_1\{s\}^{a_1}D_2\{s\}^{a_2}D_3\{s\}^{a_3}.
\]
Let us repeat the basic facts about these operators from [3].
\bLe
\label{L:Dcomm}
  The operators $D_i\{s\}$ (with the same $s$) commute with each other.
\eLe
\begin{proof}
The proof is identical to that of Lemma 3.7 in [3], using that $s$ does
not change the commutators.
\end{proof}
\bPr                                      \label{P:D-hvect}
Let $M$ be a finite-dimensional irreducible $s\ell(3)$-module
with the highest vector $m_0$ of weight $\wtt m_0=\mu$.
Any monomial $D^{a}m_0$ provides us with the highest vector in the tensor
product of $s\ell(3)$-modules $\Sym\otimes M$ that has the $s\ell(3)$-weight
given by the formula
\[
\wtt D^{\al}\{s\}m_0 = \mu + \sum a_i\, \wtt \dpind i = \mu + a_1(-1,0)+a_2(1,-1)+a_3(0,1),
\]
(whenever the expression for $\wtt D^{\al}m_0$ gives
non-dominant weight, one has
$D^{\al}m_0=0$).
Any highest weight vector in  $\Sym\otimes M$
can be written uniquely as the following linear combination
\[
w=\sum_{\al}c_{\al} D^{\al} m_0\,,\s  c_{\al}\in \CC.
\]
\ePr
We have combined here the results proven at several
places in Section 3 of [3].

In the proposition below we provide an explicit
formula for $D^{\al}\{s\}$,
but we state two lemmas first.\\

In the following $A^{[n]}:=A(A-1)\cdots(A-n+1)$.
\begin{Lemma}
  \label{L:3.3}
$(x-r)^{[m]}=\sum_{i=0}^m \,(-1)^i\,\binom{m}{i}r^{[i]}(x-i)^{[m-i]}$.
\end{Lemma}
\begin{proof}
Induction on $m$ using the fact that
\[(x-n)(x-r)^{[n]}-n((x-1)-(r-1))^{[n]} =(x-r)^{[n+1]}.\]
\end{proof}
\begin{Lemma}
  \label{L:3.4}
Let $D=ah+\partial b$ where
\begin{eqnarray*}
  \begin{array}{ll}
[a,b]=0 , & [\partial ,b] =+a , \,\, [\partial ,a] =0 \, , \\ {}
[h,b]=-2b , & [h,a]=-a , \,\, [h, \partial] =\partial  \, .
  \end{array}
\end{eqnarray*}
Then
\begin{displaymath}
  D^k = \sum^k_{m=0} \binom{k}{m}a^{k-m} \partial^m b^m (h-m)^{[k-m]} \, .
\end{displaymath}
\end{Lemma}
This is Lemma 3.8 from [3] written slightly differently.\\

Let
$ \binom{m}{i,\,j}=\frac{m!}{i!j!(m-i-j)!}. $
\begin{Proposition}
                                          \label{P:Ds-a}
Let $\al=(a,b,c)$ be a multi-index, then\\
\\
\vspace{.1cm}
$D^{\al}\{s\}=$\hspace{-2ex}
$\displaystyle{
\sum_{
\begin{array}{c}
\scriptstyle{i+j\leq a}\\
\scriptstyle{k\leq b}
\end{array}
}
}$\hspace{-2ex}
$\binom{a}{i,\,j}\binom{b}{k}
\hat{{\scriptstyle{\partial}}}
^{(a{\scriptscriptstyle{-}}i{\scriptscriptstyle{-}}j,
b{\scriptscriptstyle{+}}i{\scriptscriptstyle{-}}k,
c{\scriptscriptstyle{+}}j{\scriptscriptstyle{+}}k)}
\!\!
{\scriptstyle{
f_2^k f_1^i B^j
}}
{\scriptstyle
(h_1-i-j)^{[a{\scriptscriptstyle{-}}i{\scriptscriptstyle{-}}j]}
(h'\{s\}-j)^{[a{\scriptscriptstyle - }j]}
(h_2\{s\}-j-k)^{[b{\scriptscriptstyle{-}}k]}
}\,.$
\end{Proposition}
\begin{proof}
It is not so difficult to proceed by
induction on $\alpha$
with the help of the above lemmas.
\end{proof}
\bCo
  \label{C:3.6}
  $\ell ht \, D^{\alpha}\{s\} =\hat{\partial}^{\alpha} h_1^{[\alpha_1]}
  {h'}\{s\}^{[\alpha_1]} h_2\{s\}^{[\alpha_2]}$.
\eCo
The fact was proven in [3], but now it becomes just a corollary.
%
\begin{Proposition}
                                          \label{P:eD}
Let  ${\al=(a,b,c)}$, and
\[
{\al\!-{\!\scriptstyle{(1)}}
=(a-1,b,c)\,,\s}
{\al\!-{\!\scriptstyle{(2)}}
=(a,b-2,c)\,,\s}
{\al\!-{\!\scriptstyle{(3)}}=(a,b,c-1)\,.}
\]
Then
\begin{eqnarray*}
e_0^\pm\, D^{\al}\!\!&=&\!\! D^{\al}\{2\}\,e_0^\pm
\,\, -a\,D^{\al-(1)}\{2\}\,\dst_3^\pm B
\,\,-b\,D^{\al-(2)}\{2\}\,\dst_3^\pm f_2
\,\, -c\,D^{\al-(3)}\{1\}\,\dst_3^\pm
\\&&
+\,\,\,ab\,D^{\al-(1)-(2)}\{2\}\,\dst_3^\pm K\,,
\end{eqnarray*}
where $K=\dhind 1 f_2 h' - \dhind 2 (f_{12}h_2 - f_1f_2)$.
\end{Proposition}
Let us mention that all terms on the right are shifted by $2$ except
$cD^{\al-(3)}\{1\}\dst_3^\pm$,
where the shift is indeed $1$. 
\begin{proof}
The proof goes through the repeated use of the Newton-Jacobson formula
\[
 x\,y^k \,=\, y^k\,x +
 \sum_{i=0}^{k-1}(-1)^i\,\binom{k}{i}\,y^{k-i}(\mx{{\rm ad}}y)^i\,x
\]
and a tedious calculation of commutators. It is advisable to write
$D^\al=D_3^cD_2^bD_1^a$
making use of the commutativity of the operators. We leave the rest to
the reader.
\end{proof}
Let $C = \dhind 1 f_2 - \dhind 2 f_{12}$. Then
$ K= C h' +  \dhind 2 B $.
Let us notice that
\begin{eqnarray}
  \label{eq:Ccom}
&&C=[A,f_2]=[\,\dhind 2,B\,]\,,\notag \\
&&[\,C,B\, ]=[\,C,f_1 ]=[\,C,f_2 ]=[\,C,f_{12}]=0\,,\\
&&[\,C,\dhind i\,]=0\, \text{ for } i=1,2,3\,.\notag
\end{eqnarray}
Define
\begin{equation}
  \label{eq:Ks}
  K\{s\}=C (h'+s) +  \dhind 2 B\,.
\end{equation}
\bPr\label{P:Kmove}
\[
f_2^k f_1^i B^j\, K = K\{j+k\}f_2^k f_1^i B^j
- i\,\dhind 1 f_{12}(h'+i-1-k)f_2^k f_1^{i-1}B^j\,.
\]
\ePr
\begin{proof}
It follows from \eqref{eq:Ccom} and the following lemma.
\begin{Lemma}\label{L:Kmove}
  \begin{itemize}
  \item[(a):]$[B, K]=C B$, \s$B K\{s\}= K\{s+1\}B$, \smallskip
  \item[(b):]$[f_1,K]=-\dhind 1 f_{12}h_1$, \s
$f^i_1K\{s\} =K\{s\}f^i_1 - i\, \dhind 1 f_{12}(h_1+i-1)f^{i-1}_1$, \smallskip
  \item[(c):]$[f_2, K]=C f_2$, \s $f_2 K\{s\}= K\{s+1\}f_2$. \smallskip
  \end{itemize}
\end{Lemma}
We leave it to the reader to check the relations.
\end{proof}
\bCo \label{C:Klex}
In the notations of Proposition~\ref{P:eD}
\[  \ell ht \, D^{\al-(1)-(2)}\{2\}\,\dst_3^\pm K
=
\hat{\partial}^{\al-(2)}\dst_3^\pm f_2
h_1^{[a-1]}
  {h'}^{[a]} (h_2-1)^{[b-1]}\,.
\]
\eCo

\medskip

Following Remark~\ref{R:p-or-q} and Proposition~\ref{P:D-hvect} we
shall consider the $s\ell(3)$-highest vectors $w \in \Sym\, \La^-\La^+\, V$
in the form
\begin{equation}\label{eq:w-gen}
w=\sum_{\al} D^{\al}w_\al\,, \s w_\al \in \La^-\La^+\, V\,,
\end{equation}
and assume that
$V$ is a submodule of either  $\CC[x_1,x_2,x_3 ]$ or
$ \CC[\,\dpind1, \dpind2, \dpind3]$.
We always assume that weights $\wt w$ and $\wtt w$ are defined.

Let $N=\sdeg w$, we shall call
\begin{equation}\label{eq:w-tp}
w_{\mx{\footnotesize{top}}}=\sum_{|\si|=N} 
\,D^{\si}w_\si\,,
\end{equation}
the {\it top level} or the {\it top level terms} of $w$. We denote
\begin{equation}\label{eq:w-ntp}
w_{{\scriptscriptstyle\mx{\footnotesize{top}}-1}}=
\sum_{|\al|=N-1} 
\,D^{\al}w_\al\,,
\end{equation}
and call it the {\it near-top level terms} of $w$. Our method to
calculate $w$ will be to determine its top level first, then its
near-top level, and then the whole vector.

\bPr\label{P:e-tp}
If $e_0^-w=0$, then in the notations of \eqref{eq:w-tp}
$e_0^-w_\si=0$ for any $\si$.
\ePr
\begin{proof}
Let us use \eqref{eq:w-gen} and apply
the formula of Proposition~\ref{P:eD} in order to simplify $e_0^\pm\,w$.
It follows from the relations
\bEa\label{eq:e0m-d}
[e^-_0\,, \dst^-_i]= 0\,,\,\,\,
[e^-_0\,, \dst_1^+ ]=   \,f_2\,,\,\,
[e^-_0\,, \dst_2^+ ]= - f_{12}\,,\,\,
[e^-_0\,, \dst_3^+ ]= 0\,,
\eEa
that $e_0^- \La^-\,\La^{+}\, V \subset \La^-\,\La^{+}\, V$. This helps
to evaluate $\Sym$-degrees of the various terms, and we conclude that
\[
e_0^\pm \,w \equiv \sum_{|\si|=N} D\{2\}^{\si}e_0^\pm\, w_\si
\s\bmod \sdeg (<N)\,.
\]
The statement follows with the help of Corollary~\ref{C:3.6}.
\end{proof}
\begin{Remark}\label{R:w-tpcf}
{\sl
In particular the proposition implies that if $w$ is quasi-singular
then $w_\si$ are quasi-singular, and of course $\wt w =\wt w_\si$.
}
\end{Remark}

We need to know  explicitly the $sl(3)$-highest weight vectors in
$\La^-\,\La^{+}\, V$ for
$V\subset\CC[x_1,x_2,x_3 ]$, and $V\subset\CC[\,\dpind1, \dpind2,
\dpind3]$.
To write them down let us introduce the following notations:
\begin{eqnarray*}
            \De^{\!\!{}^\pm}(A_*)&:=& \dst^\pm_1A_1+\dst^\pm_2A_2+\dst^\pm_3A_3\,,
                                                         \qquad\qquad\qquad\\
          \hat{\De}(A_*)&:=& \dhind 1A_1+\dhind 2A_2+\dhind 3A_3\,,\\
 (A_\circ B_\circ)_{i,j}&:=&  A_iB_j -  A_jB_i \,,\\
          \cycl{a}{b}{c}&:=& a_1b_2c_3+a_2b_3c_1+a_3b_1c_2\,,\\
 \De^\pm((A_\circ B_\circ))&:=& \dst^\pm_1(A_2B_3 -  A_3B_2)+
               \dst^\pm_2(A_3B_1 -  A_1B_3) + \dst^\pm_3(A_1B_2 -  A_2B_1)\,.
\end{eqnarray*}

\subsection{\hspace*{-.8em}{\bf Semi- and quasi-singular vectors for $V\cong F(p,0)$.}}
\label{sec:3p}
We suppose that $V$ is irreducible with the highest weight $(p,0)$
throughout this subsection.
In the following proposition we list the basis of the
$s\ell(3)$-highest vectors $w$ in $\La^-\La^+V$.
We move from $\ldeg w=0$ to $\ldeg w=6$,
and listing vectors with the same $\ldeg w$ and $\wtt w$
we order them in a way that their $\wt w$ increases.
%
\bPr
\label{P:hiv-p}
Let $V$ be an irreducible $s\ell(3)$-submodule in $\CC[x_1,x_2,x_3 ]$
generated by $x_1^p$.
The $s\ell(3)$-highest vectors $w$ in $\La^-\La^+V$ are linear combination of the
following  ones
(we write $\wtt w$
at the beginning of a line):\\
$\,\ldeg w=0$
\begin{flalign*}
\s(p+0,0)&: \,x_1^p\,,&
\end{flalign*}
$\,\ldeg w=1$
\begin{flalign*}
\s(p+1,0)&:\,  \dst^-_1x_1^p\,,
            \s \dst^+_1x_1^p\,,&\\
\s(p-1,1)&:\,  (\dst^-_\circ x_\circ)_{12}\,x_1^{p-1}\,,
            \s (\dst^+_\circ x_\circ)_{12}\,x_1^{p-1}\,, &
\end{flalign*}
$\ldeg w=2$
\begin{flalign*}
\s(p+2,0)&:\,   \dst^-_1\dst^+_1x_1^p\,,&\\
\s(p+0,1)&:\,   \dst^-_{12}\,x_1^p\,,      \s
                (\dst^-_\circ\dst^+_\circ)_{12}\,x_1^{p}\,,
            \s  \dst^-_1 (\dst^+_\circ x_\circ)_{12}\,x_1^{p-1}\,,
            \s  \dst^+_{12}\,x_1^p\,,&\\
\s(p-1,0)&:\,    \cycl{\dst^-}{\dst^-}{x}x_1^{p-1}\,, \s
                 \Dem((\dst^+_\circ x_\circ))x_1^{p-1}\,,
            \s   \cycl{\dst^+}{\dst^+}{x}x_1^{p-1}\,,&\\
\s(p-2,2)&:\,    (\dst^-_\circ (\dst^+_\circ x_\circ )_{12}\, x_\circ )_{12}\,x_1^{p-2}\,,
\end{flalign*}
$\,\ldeg w=3$
\begin{flalign*}
\s(p+1,1)&:\, \dst^-_{12}\dst^+_1x_1^p\,,\s \dst^-_1\dst^+_{12}\,x_1^p\,,&\\
\s(p-1,2)&:\, \dst^-_{12}(\dst^+_\circ x_\circ )_{12}\,x_1^{p-1}\,, \s
              (\dst^-_\circ \dst^+_{12}\,x_\circ )_{12}\,x_1^{p-1}\,, &  \\
\s(p+0,0)  &:\, \dst^-_{123}\,x_1^p\,,
            \s\cycl{\dst^-}{\dst^-}{\dst^+}x_1^{p}\,,\s
              \dst^-_1\Dem((\dst^+_\circ x_\circ ))x_1^{p-1}\,,&\\
         & \s \cycl{\dst^-}{\dst^+}{\dst^+}x_1^{p}\,,\s
              \dst^-_1\cycl{\dst^+}{\dst^+}{x} x_1^{p-1}\,,\s
              \dst^+_{123}\,x_1^p\,,&\\
\s(p-2,1)&:\, \cycl{\dst^-}{\dst^-}{(\dst^+_\circ x_\circ )_{12}\,x}x_1^{p-2}\,,\s
              (\dst^-_\circ \cycl{\dst^+}{\dst^+}{x} x_\circ )_{12}\,x_1^{p-2}\,,&
\end{flalign*}
$\,\ldeg w=4$
\begin{flalign*}
\s(p+0,2)  &:\,  \dst^-_{12}\dst^+_{12}\,x_1^p\,,&\\
\s(p+1,0)&:\,  \dst^-_{123}\dst^+_1x_1^p\,,
              \s \dst^-_1\cycl{\dst^-}{\dst^+}{\dst^+}x_1^{p}\,,
            \s \dst^-_1\dst^+_{123}\,x_1^p\,,&\\
\s(p-1,1)&:\,  \dst^-_{123}(\dst^+_\circ x_\circ)_{12}\,x_1^{p-1}\,,
            \s \cycl{\dst^-}{\dst^-}{\dst^+_{12}\,x}x_1^{p-1}\,, &\\ &
            \s \dst^-_{12}\cycl{\dst^+}{\dst^+}{x}x_1^{p-1}\,,
            \s (\dst^-_\circ \dst^+_{123}\,x_\circ)_{12}\,x_1^{p-1}\,, &\\
\s(p-2,0)&:\,  \cycl{\dst^-}{\dst^-}{\cycl{\dst^+}{\dst^+}{x}x}x_1^{p-1}\,,
\end{flalign*}
$\,\ldeg w=5$
\begin{flalign*}
\s(p+0,1)&:\, \dst^-_{123}\dst^+_{12}x_1^p\,,\s\dst^-_{12}\dst^+_{123}\,x_1^p\,,&\\
\s(p-1,0)&:\, \dst^-_{123}\cycl{\dst^+}{\dst^+}{x} x_1^{p-1}\,,\s
              \cycl{\dst^-}{\dst^-}{\dst^+_{123}\,x}x_1^{p-1}\,, &
\end{flalign*}
$\,\ldeg w=6$
\begin{flalign*}
\s(p+0,0)&:\,   \dst^-_{123}\dst^+_{123}\,x_1^p\,.&
\end{flalign*}
\ePr              
\begin{proof} The statement is a generalization of Lemma 3.12 from [3].
Its proof requires rather elementary but long calculations.
We leave the details to the reader.
\end{proof}
\bCo   
\label{C:qs-deg0-p}
Admissible (i.e. of non-negative $s\ell(2)$-weight)
quasi-singular vectors
in
$\La^-\La^+V$ are:
\begin{flalign*}
\,p\geq 0&:\s   x_1^p\,,\s \dst^+_{1}\,x_1^p\,,\s \dst^-_1\dst^+_{1}\,x_1^p\,,
            \s (\dst^-_\circ \dst^+_\circ )_{12}\,x_1^{p}
             +p(\dst^-_\circ \dst^+_{1}x_\circ )_{12}\,x_1^{p-1}\,,&
\end{flalign*}
and also:
\begin{flalign*}
\,p=0&:\s \dst^+_{123}\,,\s              \dst^-_{1}\dst^+_{123}\,,\s
          \dst^-_{12}\dst^+_{123}\,,\s   \dst^-_{123}\dst^+_{123}\,.&
\end{flalign*}
\eCo
\begin{proof}
We need to apply $e^-_0$ to each of the expressions of the proposition
with non-negative $s\ell(2)$-weight
and collect the
cases when it gives zero.
It could be done easily using the relations \eqref{eq:e0m-d}.
Let us mention that later we shall utilize the results for the cases when
they are not zero.
\end{proof}

This gives us the description of admissible
quasi-singular vectors $w \in \Sym\La^-\La^+V $
such that $\sdeg w=0$. To describe the vectors with $\sdeg w>0$ is undoubtedly more
complicated, and the answer is given by the theorems below. We present the
description of
quasi-singular
vectors first,
then we use it to find
semi-singular
vectors.

\bTh
        \label{T:qs-p}
Let $V$ be an irreducible $s\ell(3)$-submodule in $\CC[x_1,x_2,x_3 ]$
generated by $x_1^p$.
The following is a complete list of admissible quasi-singular vectors
$w$
in $\Sym\La^-\La^+V$ such that
$\wt w \geq 0$ and $\sdeg w >0$, up to taking linear cominations :\\
\arabicparenlist
\begin{enumerate}
\item 
$(D^n_1\dst^-_1\dst^+_{1}-
n(p+3)D^{n-1}_1\dst^-_1(\dst^-_\circ\dst^+_\circ)_{23}\dst^+_{1}-
n^{[2]}(p+3)^{[2]}D^{n-2}_1\dst^-_{123}\dst^+_{123})x_1^p\,,\newline
\mx{$n\leq p+2$}\,,$\\
\item 
$D^n_3\dst^-_{123}\dst^+_{123}\,,\,n\geq0\,,\,p=0\,,$\\
\item 
$\hat{\De}(\dst^+_*)\hat{\De}(x_*)-
\cycl{\dst^-}{\dst^+}{\dst^+}\hat{\De}(x_*)+
\hat{\De}(\dst^-_*)\cycl{\dst^+}{\dst^+}{x}+
\cycl{\dst^-}{\dst^-}{\dst^+_{123}\,x}\,,\,p=1\,,$\\
\item 
$D_2\!\left((\dst^-_\circ \dst^+_\circ )_{12}\,x_1^{p}
        +p(\dst^-_\circ \dst^+_{1}x_\circ )_{12}\,x_1^{p-1}\right)
           -\dst^-_1(\dst^-_\circ\dst^+_\circ)_{23}\,{\dst^+_1}x_1^{p}\,,\,p\geq0\,,$\\
\item 
$w=\hat{\De}(x_*)-\tfrac{1}{2}\Dem((\dst^+_\circ x_\circ ))\,,\,p=1\,,$\\
\item 
$\tfrac{1}{3}D_1\dst^+_{1}x_1-\cycl{\dst^-}{\dst^+}{\dst^+}x_1+
       \dst^-_1\cycl{\dst^+}{\dst^+}{x}\,,\,p=1\,,$\\
\item 
$D_1\!\left(\dst^-_1f_1(\dst^+_{1}x_1) - 2\dst^-_2 \dst^+_{1}x_1\right)
   -3\left(\cycl{\dst^-}{\dst^-}{\dst^+_{12}\,x} + \dst^-_{12}\cycl{\dst^+}{\dst^+}{x}\right)\,,\,p=1 \,,$\\
\item 
$\hat{\De}(\dst^+_*) - \cycl{\dst^-}{\dst^+}{\dst^+}\,,\,p=0\,,$\\
\item 
$\hat{\De}(\dst^-_*)\dst^+_{123}\,,\,p=0\,,$\\
\item 
$(\dhind 2\dst^-_{12}+\dhind 3\dst^-_{13})\dst^+_{123}\,,\,p=0\,.$\\
\end{enumerate}
%
\eTh
Let us notice that only in (1) $p$ and $\sdeg w$ are unbounded,
for all cases, except (1) and (2), we have $\sdeg w \leq 2$,
and for all cases, except (1) and (4), $p\leq 1$.\\

We shall check first that all these vectors are indeed
quasi-singular. While checking (1) we
use Proposition~\ref{P:eD}.
To check the rest becomes elementary calculation
based on the relations \eqref{eq:e0m-d} and the fact that
\begin{equation}
  \label{eq:e-dpart}
[e^\pm_0\,, \dhind i]= 0\,.\,\,\,
\end{equation}

\bTh
      \label{T:ss-p}
Let $V$ be an irreducible $s\ell(3)$-submodule in $\CC[x_1,x_2,x_3 ]$
generated by $x_1^p$.
The admissible semi-singular vectors $w$
in $\Sym\La^-\La^+V$ such that
$\wt w \geq 0$ and $\sdeg w >0$ exist only when $p=0$ and they
are the following:\\
\romanparenlist
\begin{enumerate}
\item 
$\s \hat{\De}(\dst^+_*) - \cycl{\dst^-}{\dst^+}{\dst^+}\,,$\\
\item 
$\s  \hat{\De}(\dst^-_*)\,(\hat{\De}(\dst^+_*)
      - \cycl{\dst^-}{\dst^+}{\dst^+})
\,,$\\
\item 
$\s \dst^-_{1}\hat{\De}(\dst^+_*)
       - \dst^-_{1}(\dst^-_\circ\dst^+_\circ)_{23}\dst^+_{1}=
\dst^-_{1}(\hat{\De}(\dst^+_*) - \cycl{\dst^-}{\dst^+}{\dst^+})
\,,$\\
\item 
$\s \hat{\De}(\dst^-_*)\,\dst^+_{123}\,,$\\
\item 
$\s (\dhind 2\dst^-_{12}+\dhind 3\dst^-_{13})\dst^+_{123}=
\dst^-_{1}\hat{\De}(\dst^-_*)\,\dst^+_{123}
\,.$\\

\end{enumerate}
\eTh

We also need to check that vectors listed in Theorem~\ref{T:ss-p} are semi-singular.
Here the calculations use  the relations of
commutation with $e_0^+$:
\bEa\label{eq:e0p-d}
[e^+_0\,, \dst_1^- ]= -  \,f_2\,,\,\,
[e^+_0\,, \dst_2^- ]=  f_{12}\,,\,\,
[e^+_0\,, \dst_3^- ]= 0\,.\,\,\,
[e^+_0\,, \dst^+_i]= 0\,,\,\,\,
\eEa
and are pretty straightforward.

What left is the hard parts of the theorems,  the statements that
the lists contain all the vectors. The proof of this demands to work through
an elaborated ``tree of cases'' and will be postponed to the next section.

\subsection{\hspace*{-.8em}{\bf Semi- and quasi-singular vectors for
    $V\cong F(0,q)$.}}
\label{sec:3q}
We suppose that $V$ is irreducible with the highest weight $(0,q)$,
$q\geq 1 $
throughout this subsection.

Similarly we need first of all the description of the
$s\ell(3)$-highest vectors $w$ in $\La^-\La^+V$.

\bPr
\label{P:hiv-q}
Let $V$ be an irreducible $s\ell(3)$-submodule in
$\CC[\dpind 1,\dpind 2,\dpind 3 ]$ generated by $\pt_3^q$.
The $s\ell(3)$-highest vectors $w$ in $\La^-\La^+V$ are linear combination of the
following ones:\\
$\,\ldeg w=0$
\begin{flalign*}
\s(0,q+0)&: \,\pt_3^q\,,&
\end{flalign*}
$\,\ldeg w=1$
\begin{flalign*}
\s(0,q-1)&: \, \Dem(\pt_*)\pt_3^{q-1}\,, \s \Dep(\pt_*)\pt_3^{q-1}\,,&\\
\s(1,q+0)&: \, \dst^-_1\pt_3^q\,,         \s \dst^+_1\pt_3^q\,,&
\end{flalign*}
$\,\ldeg w=2$
\begin{flalign*}
\s(0,q-2)&: \, \Dem(\Dep(\pt_*)\pt_*)\pt_3^{q-2}\,,&\\
\s(1,q-1)&: \, \dst^-_1\Dem(\pt_*)\pt_3^{q-1}\,, \s
               \dst^-_1\Dep(\pt_*)\pt_3^{q-1}\,, \s
               \Dem(\dst^+_1\pt_*)\pt_3^{q-1}\,, \s
               \dst^+_1\Dep(\pt_*)\pt_3^{q-1}\,, &\\
\s(0,q+1)&: \, \dst^-_{12}\pt_3^q\,,      \s
               (\dst^-_\circ \dst^+_\circ )_{12}\pt_3^q\,,      \s
               \dst^+_{12}\pt_3^q \,,&\\
\s(2,q+0)&: \, \dst^-_1\dst^+_1\pt_3^q\,, &
\end{flalign*}
$\,\ldeg w=3$
\begin{flalign*}
\s(1,q-2)&: \, \dst^-_1\Dem(\Dep(\pt_*)\pt_*)\pt_3^{q-2}\,, \s
               \Dem(\dst^+_1\Dep(\pt_*)\pt_*)\pt_3^{q-2}\,,&\\
\s(0,q+0)&: \, \dst^-_{123}\pt_3^q\,, \s \dst^-_{12}\Dep(\pt_*)\pt_3^{q-1}\,,\s
               \cycl{\dst^-}{\dst^-}{\dst^+}\pt_3^q\,, &\\&
             \s
               (\dst^-_\circ \dst^+_\circ )_{12}\Dep(\pt_*)\pt_3^{q-1}\,, \s
               \cycl{\dst^-}{\dst^+}{\dst^+} \pt_3^q\,, \s
               \dst^+_{123}\pt_3^q\,,&\\
\s(2,q-1)&: \, \dst^-_1\Dem(\dst^+_1\pt_*)\pt_3^{q-1}\,, \s
               \dst^-_1 \dst^+_1\Dep(\pt_*)\pt_3^{q-1}\,, &\\
\s(1,q+1)&: \,  \dst^-_{12}\dst^+_1\pt_3^q\,,  \s \dst^-_1\dst^+_{12}\pt_3^q\,, &
\end{flalign*}
$\,\ldeg w=4$
\begin{flalign*}
\s(0,q-1)&: \,  \dst^-_{123}\Dep(\pt_*)\pt_3^{q-1}\,, \s
                \cycl{\dst^-}{\dst^-}{\dst^+}\Dep(\pt_*)\pt_3^{q-1}\,,\s
                \Dem(\dst^+_{123}\pt_*)\pt_3^{q-1}\,,                &\\
\s(2,q-2)&: \,  \dst^-_1\Dem(\dst^+_1\Dep(\pt_*)\pt_*)\pt_3^{q-2}\,, &\\
\s(1,q+0)&: \,  \dst^-_{123}\dst^+_1\pt_3^q\,,  \s
                \dst^-_{12}\dst^+_1\Dep(\pt_*)\pt_3^{q-1}\,, \s
                \dst^-_1\cycl{\dst^-}{\dst^+}{\dst^+} \pt_3^q\,, \s
                \dst^-_1\dst^+_{123}\pt_3^q\,,    &\\
\s(0,q+2)&: \,  \dst^-_{12}\dst^+_{12}\pt_3^q\,,  &
\end{flalign*}
$\,\ldeg w=5$
\begin{flalign*}
\s(1,q-1)&: \, \dst^-_{123}\dst^+_1\Dep(\pt_*)\pt_3^{q-1}\,, \s
               \dst^-_1\Dem(\dst^+_{123}\pt_*)\pt_3^{q-1}\,, &\\
\s(0,q+1)&: \, \dst^-_{123}\dst^+_{12}\pt_3^q\,,  \s \dst^-_{12}\dst^+_{123}\pt_3^q\,, &
\end{flalign*}
$\,\ldeg w=6$
\begin{flalign*}
\s(0,q+0)&: \, \dst^-_{123}\dst^+_{123}\pt_3^q\,. &
\end{flalign*}
\ePr
\begin{proof} This is also a generalization of Lemma 3.12 from [3]
and we leave it as an exercise to the reader.
\end{proof}
\bCo \label{C:qs-deg0-q}
Admissible
quasi-singular vectors
in
$\La^-\La^+V$ are:
\begin{flalign*}
&\s \pt_3^q\,, \s \Dep(\pt_*)\pt_3^{q-1}\,,
 \s   \dst^-_1\Dep(\pt_*)\pt_3^{q-1}\,,\s \text{ where } q\geq 1 \,,\s
\Dem(\Dep(\pt_*)\pt_*)\pt_3^{q-2}\,, \s q\geq 2
\,,&
\end{flalign*}
and more for $q=1$:
\begin{flalign*}
&\s \dst^+_1\Dep(\pt_*)\,, \s
(\dst^-_\circ \dst^+_\circ )_{12}\Dep(\pt_*)\,, \s
    \dst^-_1\dst^+_1\Dep(\pt_*)\,, \s
    \cycl{\dst^-}{\dst^-}{\dst^+}\Dep(\pt_*)\,, \s
    \dst^-_{12}\dst^+_1\Dep(\pt_*)\,.&
\end{flalign*}
\eCo
We need only to calculate the action of $e_0^-$ on the
vectors of non-negative $s\ell(2)$-weight
of the proposition. Corollary collects the cases when the results are zero.
The results for the cases when the results
are non-zero will be needed later too. \\

Admissible semi-singular and quasi-singular vectors $w$ with $\sdeg w \geq 0$
are described by the following theorems.
%
\bTh
\label{T:qs-q}
Let $V$ be an irreducible $s\ell(3)$-submodule in $\CC[\,\pt_1,\pt_2,\pt_3 ]$
generated by $\pt_3^q$, $q\geq 1$.
The admissible quasi-singular vectors $w$
in $\Sym\La^{\!-}\La^{\!+}V$ such that
$\wt w \geq 0$ and $\sdeg w >0$ are given by the following expressions:\\
\arabicparenlist
\begin{enumerate}
\item 
$D^n_3\Dem(\Dep(\pt_*)\pt_*)\pt_3^{q-2}
+\frac{n}{q-1}
D^{n-1}_3\cycl{\dst^-}{\dst^-}{\dst^+}\Dep(\pt_*)\pt_3^{q-1}
-\frac{n^{[2]}}{q^{[2]}}
D^{n-2}_3\dst^-_{123}\dst^+_{123}\pt_3^q
\,,\newline
n\geq 1,\,\,q\geq 2\,,
$\\
\item 
$D_1\,\dst^-_1\!\Dep(\pt_*)\pt_3^{q-1}
+\cycl{\dst^-}{\dst^-}{\dst^+}\Dep(\pt_*\!)\pt_3^{q-1}
\,,
$\\
\item 
 $D_1D_2\,\dst^-_1\!\Dep\!(\pt_*\!)\pt_3
 -\tfrac{1}{2}D_1\dst^-_1\!\Dem(\dst^+_1\!\Dep(\pt_*\!)\pt_*\!)
 -\tfrac{3}{2}D_2\cycl{\dst^-}{\dst^-}{\dst^+}\Dep(\pt_*\!)\pt_3\,,
\,\, q=2\,,
$\\
\item 
$D_2\,\dst^-_1\!\Dep(\pt_*)\pt_3 -
\dst^-_1\!\Dem(\dst^+_1\!\Dep(\pt_*\!)\pt_*\!)   \,,\,\,q=2\,,
$\\
\item 
$D_2\pt_3 - \Dem(\dst^+_1\pt_*) \,,\,\,q=1\,,$\\
\item 
$D_1\dst^+_1\Dep(\pt_*)-2\Dem(\dst^+_{123}\pt_*) \,,\,\,q=1\,,
$\\
\item 
$D_2(\dst^-_\circ \dst^+_\circ )_{12}\Dep(\pt_*)
-\dst^-_1\Dem(\dst^+_{123}\pt_*)\,,\,\,q=1\,,
$\\
\item 
$D_3^n\cycl{\dst^-}{\dst^-}{\dst^+}\Dep(\pt_*)
-nD_3^{n-1}\dst^-_{123}\dst^+_{123}\pt_3\,, \,\, n \geq 1\,,\,\,q=1\,,$\\
\item 
$D_1\dst^-_{12}\dst^+_1\Dep(\pt_*)
-2\,\dst^-_{123}\dst^+_{123}\pt_3\,,\,\,q=1\,,$\\
\item 
$D_1^n\,\dst^-_1\dst^+_1\Dep(\pt_*)
-3nD_1^{n-1}\dst^-_1\Dem(\dst^+_{123}\pt_*), \,\,n=1,2\,,\,\,q=1
\,.$\\
\end{enumerate}
This is a complete list of such vectors up to taking linear combinations.
\eTh
It is not difficult to check that these vectors are indeed
quasi-singular. For example for the case (3), after
making calculations according to Proposition~\ref{P:eD} we get
\begin{eqnarray*}
e_0^-D_1D_2\,\dst^-_1\!\Dep\!(\pt_*\!)\pt_3 &=&
+D_2\{2\}\dst^-_{3}\Dem(\!\Dep(\pt_*\!)\pt_*\!)\notag\\
&&
+D_1\{2\}\dst^-_{31}\Dep(\pt_*\!)\pt_2 \\
&&+\dhind 1(-3\,\dst^-_{31}\Dep(\pt_*\!)\pt_2)
+\dhind 2( \,2\,\dst^-_{31}\Dep(\pt_*\!)\pt_1
- \dst^-_{32}\Dep(\pt_*\!)\pt_2)\,. \notag
\end{eqnarray*}
At the same time
\begin{eqnarray*}
D_1\{2\}\dst^-_{31}\Dep(\pt_*\!)\pt_2&=&
\,\dhind 1(+6\,\dst^-_{31}\Dep(\pt_*\!)\pt_2)
\,\,+\dhind 2(-3\,\dst^-_{31}\Dep(\pt_*\!)\pt_1
+3\, \dst^-_{32}\Dep(\pt_*\!)\pt_2)\,,\notag\\
D_2\{2\}\dst^-_{3}\Dem(\!\Dep(\pt_*\!)\pt_*\!)&=&
\qquad\qquad\qquad\qquad\qquad
\dhind 2(+ \,\dst^-_{31}\Dep(\pt_*\!)\pt_1
+ \dst^-_{32}\Dep(\pt_*\!)\pt_2)\,.\notag
\end{eqnarray*}
Also
\begin{equation*}
e_0^- D_1\dst^-_1\Dem(\dst^+_1\!\Dep(\pt_*\!)\pt_*\!)=
D_1\{2\}\dst^-_{31}\Dep(\pt_*\!)\pt_2
-3\,\dst^-_{312}\dst^+_{3}\Dep(\pt_*\!)\pt_2\,,
\end{equation*}
and
\begin{equation*}
e_0^- D_2\cycl{\dst^-}{\dst^-}{\dst^+}\Dep(\pt_*\!)\pt_3
=
D_2\{2\}\dst^-_{3}\Dem(\!\Dep(\pt_*\!)\pt_*\!)
+\dst^-_{312}\dst^+_{3}\Dep(\pt_*\!)\pt_2\,.
\end{equation*}
Therefore we conclude that
\[
e_0^-\left(
D_1D_2\,\dst^-_1\Dep\!(\pt_*\!)\pt_3
-\tfrac{1}{2}D_1\dst^-_1\Dem(\dst^+_1\!\Dep(\pt_*\!)\pt_*\!)
-\tfrac{3}{2}D_2\cycl{\dst^-}{\dst^-}{\dst^+}\Dep(\pt_*\!)\pt_3\right) =0
\,.
\]
We leave to check the rest of vectors to the reader. The proof that
there are no other quasi-singular vectors will be presented later in a
separate section.

\bTh
      \label{T:ss-q}
Let $V$ be irreducible $s\ell(3)$-submodule in $\CC[\pt_1,\pt_2,\pt_3 ]$
generated by $\pt_3^q$, $q\geq 1$.
The admissible semi-singular vectors $w$
in $\Sym\La^{\!-}\La^{\!+}V$ such that
$\wt w \geq 0$ and $\sdeg w >0$ exist only for $q=1$ and $q=2$. They
are the following vectors:\\
\romanparenlist
\begin{enumerate}
\item 
$\s \dhind 2\pt_3-\dhind 3\pt_2 - \Dem(\dst^+_1\pt_*) \,,$\\
\item 
$\s \dhind 2\dst^-_1\Dep(\pt_*)\pt_3
-\dhind 3\dst^-_1\Dep(\pt_*)\pt_2 - \dst^-_1\Dem(\dst^+_1\Dep(\pt_*)\pt_*)   \,,$\\
\item 
$\s (\hat{\De}(\dst^-_*) - \cycl{\dst^-}{\dst^-}{\dst^+})\Dep(\pt_*\!)\,,$\\
\item 
$\s (\dhind 1 \dst^+_1 +\dhind 2 \dst^+_2 +
\dhind 3 \dst^+_3)\Dep(\pt_*)-\Dem(\dst^+_{123}\pt_*\!)\,,$\\
\item 
$\s (\dhind 1 \dst^-_1\dst^+_1 +\dhind 2 \dst^-_1\dst^+_2 +
\dhind 3 \dst^-_1\dst^+_3)\Dep(\pt_*)- \dst^-_1\Dem(\dst^+_{123}\pt_*\!)\,,$\\
\item 
$\s D_1^2\,\dst^-_1\dst^+_1\Dep(\pt_*)
-6\,D_1\,\dst^-_1\Dem(\dst^+_{123}\pt_*)\,.$\\
\end{enumerate}
\eTh
It is not so difficult
to check that these vectors are semi-singular.
The proof that Theorems~\ref{T:ss-p}, \ref{T:ss-q} describe all such
vectors will be discussed later in a separate section.

\bCo \label{C:ss-deg0-q}
Admissible semi-singular vectors
with
\mbox{$\sdeg w=0$} in
$\La^-\La^+V$ are:
\begin{flalign*}
&\s \pt_3^q\,, \s \Dep(\pt_*)\pt_3^{q-1}\,,
 \s  \text{ where } q\geq 1 \,,\s
\Dem(\Dep(\pt_*)\pt_*)\,, \s q= 2
\,,&
\end{flalign*}
\begin{flalign*}
&\s \dst^+_1\Dep(\pt_*)\,, \s \dst^-_1\Dep(\pt_*)\,,\s
    \dst^-_1\dst^+_1\Dep(\pt_*)
\,, \s q=1\,.&
\end{flalign*}
\eCo
One easily gets the list from Corollary~\ref{C:qs-deg0-q} applying
$e_0^+$.  We shall use it in Section~\ref{sec:8}.


\section{Proof of Theorem~\ref{T:qs-p}.}
\label{sec:4}

We consider a quasi-singular vector $w \in \Sym\La^-\La^+V$ under the
conditions of Theorem~\ref{T:qs-p}. Without loss of generality we can
assume that $w$ is a $g\ell(1)$-weight vector.
We keep the notations of \eqref{eq:w-gen} and
\eqref{eq:w-tp}. \\

First of all let us notice that for $|\al|=N-i$
\begin{equation}\label{eq:4.1}
\ldeg w_\al = \ldeg w_\ka + 2i \,,\s \wt w_\al=\wt w_\ka=\wt w\,,
\end{equation}
because $\wt w=\wt D^\al w_\al = \wt w_\al$, and
$\wtg w=\wtg D^\al w_\al = -\tfrac{2}{3}|\al| + \wtg w_\al$,
but $\wtg$ of an element in $\La^-\La^+V$ depends only on its $\La$-degree.

Suppose that $\ka$ is the lexicographically highest element in
\[
I_{\text{top}}:=\{\si\,|\,w_\si\neq 0\,, \,\,
\text{ where } |\si|=N=\sdeg w \,\}\,,
\]
(i.e. $w_\si$ are the top level coefficients  of $w$).
It is clear that in order to prove that a quasi-singular vector
is a linear combination of vectors listed in the statement of the theorem,
it is enough to show that the list contains a vector for each
possible ``highest term'' $w_\ka$, then the result follows by
induction. Therefore it is enough to prove that any possible ``highest
term'' $w_\ka$ is present among the ``highest terms'' $w_\ka$ of the
vectors listed in Theorem~\ref{T:qs-p}.

\begin{Remark}\label{R:4.1}
{\sl
It follows from \eqref{eq:4.1} that
 $\ldeg  w_\si=\ldeg w_\ka$ and $\wt w_\si=\wt w_\ka$.
These are important restrictions on $w_\si$.
}
\end{Remark}

Proposition~\ref{P:e-tp} and  Remark~\ref{R:w-tpcf}
show that $w_\si$ are quasi-singular vectors of non-negative $s\ell(2)$-weight,
thus belong to the list given by Corollary~\ref{C:qs-deg0-p}.
\bPr\label{P:4.2}

{\rm (1):}
If
$\,\,w_\ka = x_1^p\,, \s\dst_1^-x_1^p\,, $ or $
\mx{ $(\dst^-_\circ \dst^+_\circ )_{12}\,x_1^{p}
             +p(\dst^-_\circ \dst^+_{1}x_\circ )_{12}\,x_1^{p-1}\,$},$
then $\,I_{\text{top}}=\{ \ka \}$.\newline
{\rm (2):}
If $\,\,w_\ka =\dst_1^-\dst_1^+x_1^p$,
then either $\,I_{\text{top}}=\{ \ka \}$ or
$\,I_{\text{top}}=\{ \ka, \si \}$,
where $\si=\ka -(1,-1,0)$
and $\,w_\si=c_\si((\dst^-_\circ \dst^+_\circ )_{12}\,x_1^{p}
             +p(\dst^-_\circ \dst^+_{1}x_\circ )_{12}\,x_1^{p-1})\,,
   \,c_\si\in \CC $.\newline
{\rm (3):}
If
$\,\,
w_\ka =\dst_{123}^+\,,
\s\dst_{1}^-\dst_{123}^+\,, \s\dst_{12}^-\dst_{123}^+\,, \text{ or }
\,\dst_{123}^-\dst_{123}^+\,,
$
then $\,\,I_{\text{top}}=\{ \ka \}$.
\ePr

\begin{proof}
The result follows from Remark~\ref{R:4.1},  Remark~\ref{R:w-tpcf} and
the fact that  $\wtt D^\si w_\si = \wtt D^\ka w_\ka$ as soon as one
checks the weights $\wtg$,  $\wt$, $\wtt$
through the list of Corollary~\ref{C:qs-deg0-p}.
\end{proof}

The proposition describes possibilities for
$w_{\mx{\footnotesize{top}}}$,
now
we are to determine what is possible for
$w_{{\scriptscriptstyle\mx{\footnotesize{top}}-1}}$.
We are interested only in terms $D^\al w_\al \neq 0$ with $|\al|=N-1$
and $e_0^-w_\al\neq 0$.
The following will be of use.
\begin{Remark}\label{R:4.3}
{\sl
Let $ D^\al w_\al\neq 0$, and consider any $\be$, such that
$0 \leq \be_i \leq \al_i$. Then $D^\be w_\al\neq 0$, because
$D^\al w_\al=D^{\al-\be} D^\be w_\al$, hence
the weight
\[
\wtt D^\be w_\al =\wtt w_\al + \be_1(-1,0)+\be_2(1,-1)+\be_3(0,1)
\]
is dominant by Proposition~\ref{P:D-hvect}. For example one can take
$\be=(\al_1,0,0)$ or $\be=(0,\al_2,0)$.
}
\end{Remark}
By Proposition~\ref{P:eD} the condition $e_0^-w=0$ implies
\begin{equation}
             \label{eq:4.2}
\sum_{|\al|=N-1}D\{2\}^\al e_0^- w_\al
\equiv  - e_0^-w_{\mx{\footnotesize{top}}}
\s \bmod \,\sdeg(<N-1)\,.
\end{equation}
Applying again Proposition~\ref{P:eD} we conclude that
\begin{equation}
  \label{eq:4.4}
\sum_{|\al|=N-1}D\{2\}^\al e_0^- w_\al
\equiv
\ka_3\,D^{\ka-(3)}\{1\}\,\dst_3^- w_\ka
\s \bmod \,\lxl{\,\ka{\scriptstyle{-(3)\,}}}\,,
\end{equation}
because all the other terms in $e_0^-w_{\mx{\footnotesize{top}}}$
are lexicographically smaller. From
%
%
%
now on we study the possibilities of Proposition~\ref{P:4.2}
{\bf case by case}.\\

{\bf Case} 1(i): $\underline{w_\ka=x_1^p}$.
\smallskip\\
Here Remark~\ref{R:4.3} shows that $\ka_1\leq p$ and $\ka_2=0$.
\bLe $\ka_3=0$.
\eLe
\begin{proof}


In our case $w_{\mx{\footnotesize{top}}}=D^\ka x_1^p$. From
\eqref{eq:4.4} it follows that
either
$\ka_3=0$ or $e_0^-w_{\ka{\scriptstyle{-(3)\,}}}\neq 0$.
We shall show that the latter is impossible.

We are to look for the values of $w_{\ka{\scriptstyle{-(3)\,}}}$.
First of all, from
\eqref{eq:4.1} we see that
$\wt w_{\ka{\scriptstyle{-(3)\,}}}=0$,
$\ldeg w_{\ka{\scriptstyle{-(3)\,}}} =2$, thus
$w_{\ka{\scriptstyle{-(3)\,}}}\in\La^-_1\La^+_1V$.
As $\wtt D^\ka w_\ka =\wtt D^\al w_\al$,
it is clear that $\wtt w_{\ka{\scriptstyle{-(3)\,}}}= (p,1)$,
and the list of Proposition~\ref{P:hiv-p} provides us with the only
choice
\[
w_{\ka{\scriptstyle{-(3)\,}}}= c_1
    (\dst^-_\circ\dst^+_\circ)_{12}\,x_1^{p}\,+
            c_2 \, \dst^-_1 (\dst^+_\circ x_\circ)_{12}\,x_1^{p-1}\,.
\]
Then $e_0^-w_{\ka{\scriptstyle{-(3)\,}}}= \dst_1^- A+\dst_2^-B$
for some $A,B\in \La^+V$ because $[\,e_0^-,\,\dst^-_i]=0$, and
the equality \eqref{eq:4.4} is impossible when
$e_0^-w_{\ka{\scriptstyle{-(3)\,}}}\neq 0$.
\end{proof}

We conclude that $\ka=(N,0,0)$, and \eqref{eq:4.2}, \eqref{eq:4.4}
give us
\begin{equation}
  \label{eq:4.5}
\sum_{|\al|=N-1}D\{2\}^\al e_0^- w_\al
\equiv
\ka_1\,D^{\ka-(1)}\{2\}\,\dst_3^-B\, x_1^p
\s \bmod \,\lxl{\,\ka{\scriptstyle{-(1)\,}}}\,,
\end{equation}
where $B\,x_1^p = (p+1)\,x_3x_1^{p-1}$.
\bLe $N=1$, $p=1$.
\eLe
\begin{proof}
By the same arguments as above $w_{\ka{\scriptstyle{-(1)\,}}}\in\La^-_1\La^+_1V$,
$\wt w_{\ka{\scriptstyle{-(1)\,}}}=0$, and it is easy to see that
$\wtt w_{\ka{\scriptstyle{-(1)\,}}}= (p-1,0,0)$.
Then from Proposition~\ref{P:hiv-p} it follows that
$ w_{\ka{\scriptstyle{-(1)\,}}}=
c\, \Dem((\dst^+_\circ x_\circ))x_1^{p-1}$.
We calculate that
\[
e_0^-w_{\ka{\scriptstyle{-(1)\,}}}=c\left(
(p-1)\dst_1^-x_3^2x_1^{p-2}-(p+1)\dst_3^-x_3x_1^{p-1}\right)\,.
\]
Clearly \eqref{eq:4.5} implies that $p=1$. But $N=\ka_1\leq p$ in our case,
and because $N=\sdeg w >0$ we get $N=1$.
\end{proof}
As a result we have come to the vector given in Theorem~\ref{T:qs-p}(5).\\

{\bf Case} 1(ii): $\underline{w_\ka=\dst_1^+\,x_1^p}$.\\
We follow the same line of arguments.
By  Remark~\ref{R:4.3}
$\ka_1\leq p+1$  and  $\ka_2=0$ in this case.

\bLe $\ka_3=0$.
\eLe
\begin{proof}
Consider the terms $D^\al w_\al$ with $|\al|=N-1$
and $e_0^-w_\al\neq 0$. Here we immediately get
$\wt w_\al=+1$ and
$w_\al\in\La^-_1\La^+_2 V$.
Now
from \eqref{eq:4.4} it follows that
if $\ka_3\neq 0$, then $e_0^-w_{\ka{\scriptstyle{-(3)\,}}}\neq 0$, and we
are to look for the value of $w_{\ka{\scriptstyle{-(3)\,}}}$ in the
list of Proposition~\ref{P:hiv-q}. In this case
$\wtt w_{\ka{\scriptstyle{-(3)\,}}}= (p+1,1,0)$ and we get
$w_{\ka{\scriptstyle{-(3)\,}}}=
c\,\dst^-_1\dst^+_{12}\,x_1^p\,$.
But this does not fit into \eqref{eq:4.4}.
\end{proof}

We come to $\ka=(N,0,0)$, and to the equation
\begin{equation}
  \label{eq:4.8}
\sum_{|\al|=N-1}D\{2\}^\al e_0^- w_\al
\equiv
\ka_1\,D^{\ka-(1)}\{2\}\,\dst_3^-B\,\dst_1^+\, x_1^p
\s \bmod \,\lxl{\,\ka{\scriptstyle{-(1)\,}}}\,,
\end{equation}
where $B\,\dst_1^+\,x_1^p = (p+2)\,f_{12}\dst_1^+\,x_1^{p-1}$.
We also see that
$\wtt w_{\ka{\scriptstyle{-(1)\,}}}= (p,0,0)$.
Then from Proposition~\ref{P:hiv-p} it follows that
$ w_{\ka{\scriptstyle{-(1)\,}}}=
c_1\, \cycl{\dst^-}{\dst^+}{\dst^+}x_1^{p}+
            c_2\,  \dst^-_1\cycl{\dst^+}{\dst^+}{x} x_1^{p-1}$.
Hence
\begin{eqnarray*}
e_0^-  w_{\ka{\scriptstyle{-(1)\,}}}=&&
c_1\,(p\,\dst_1^-\dst_3^+\,x_3x_1^{p-1}-
\dst_3^-\dst_3^+\,x_1^{p}-
p\dst_3^-\dst_1^+\,x_3x_1^{p-1})\\
&+&
c_2\,(p\,\dst_1^-\dst_3^+\,x_3x_1^{p-1}-(p-1)
\,\dst_1^-\dst_3^+\,x_3^2x_1^{p-2})\,.
\end{eqnarray*}
Clearly $p\leq 1$ and either $p=1,\,c_1+c_2=0$, or
$p=0,\, c_2=0$ could satisfy \eqref{eq:4.8}.
Now $p=1,\,N=1$ gives us Theorem~\ref{T:qs-p}(6),
$p=1,\,N=2$ gives Theorem~\ref{T:qs-p}(3) and
$p=0,\,\,N=1$ gives Theorem~\ref{T:qs-p}(8).\\

{\bf Case} 1(iii): $\underline{
w_\ka=\mx{ $(\dst^-_\circ \dst^+_\circ )_{12}\,x_1^{p}
     +p(\dst^-_\circ \dst^+_{1}x_\circ )_{12}\,x_1^{p-1}\,$}.}$\\
By  Remark~\ref{R:4.3} we see that
$\ka_1\leq p$  and  $\ka_2\leq 1$.

\bLe
$\ka_3=0$.
\eLe
\begin{proof}
Here
$\wt w_{\al}=0$,
$\ldeg w_{\al} =4$ for $|\al|=N-1$,
hence
$w_{\al}\in\La^-_2\La^+_2 V$.

For $\al={\ka{\scriptstyle{-(3)\,}}}$ we have
$\wtt  w_{\ka{\scriptstyle{-(3)\,}}}= \wtt w_\al + (0,1) = (p,2)$.
Therefore we find with the help of Proposition~\ref{P:hiv-p} that
$w_{\ka{\scriptstyle{-(3)\,}}}= c\,\dst_{12}^-\dst_{12}^+\,x_1^p$.
But from \eqref{eq:4.4} it follows that
\begin{equation}
  \label{eq:4.10}
\sum_{|\al|=N-1}D\{2\}^\al e_0^- w_\al
\equiv
\ka_3\,D^{\ka-(3)}\{1\}\,\dst_3^-\dst_1^-w_\ka
\s \bmod \,\lxl{\,\ka{\scriptstyle{-(3)\,}}},
\end{equation}
which is only possible with the above value of $w_{\ka{\scriptstyle{-(3)\,}}}$
when $\ka_3= 0$, $c=0$.
\end{proof}
Let us remind that $\ka_2\leq 1$.

{\bf Subcase}: $\ka_2= 1$.\\
Let $\ka=(n,1,0)$.
Now from Proposition~\ref{P:eD} we get
\begin{equation}
  \label{eq:4.11}
e_0^-D^\ka w_\ka = D_1^{n-1}\{2\}\left(
        -n\, D_2\{2\}\dst_3^-B\,w_\ka
                  - D_1\{2\}\dst_3^-f_2 w_\ka
                          +n\,\dst_3^-K\,w_\ka
\right)\,.
\end{equation}
This and \eqref{eq:4.2} implies
\begin{equation}\notag
D_1^{n-1}\{2\}\left(
D_2\{2\}\dst_3^-B +
                  n( D_1\{2\}\dst_3^-f_2
- \dst_3^-K )
\right) w_\ka
\equiv  \sum_{|\al|=n}D\{2\}^\al e_0^- w_\al
\bmod  \sdeg(< n)\,,
\end{equation}
where we will ignore those $\al$ where $e_0^- w_\al=0$.

Let us consider the weaker congruence,
the same expression but
\[ \bmod \, \dlxl{(n-1,1,0)},\]
i.e. we use the degree-lexicographic order defined in
\eqref{eq:3.1}. Then
only $\al={(n,0,0)}$ and $\al={(n-1,1,0)}$
are to be considered, and we come to an expression
\begin{eqnarray}\notag
&D_1^{n-1}\{2\}\left(\,
 D_1\{2\}\dst_3^-f_2 + n ( D_2\{2\}\dst_3^-B - \dst_3^-K )
                     \right)w_\ka
                               \equiv &\\
& \s \equiv D_1^{n-1}\{2\}\left(
D_1\{2\}e_0^- w_{(n,0,0)}+ D_2\{2\}e_0^- w_{(n-1,1,0)}
\right)&
\bmod \, \dlxl{{\scriptstyle(n-1,1,0)}}\,.\notag
\end{eqnarray}
It is not difficult to prove that this implies the following
congruence
\begin{eqnarray}\notag
&(
 D_1\{2\}\dst_3^-f_2 + n ( D_2\{2\}\dst_3^-B - \dst_3^-K )
 \,)w_\ka \equiv & \\
&\s \equiv D_1\{2\}e_0^- w_{(n,0,0)}+ D_2\{2\}e_0^- w_{(n-1,1,0)}
&
\bmod \, \dlxq{{\scriptstyle(0,0,1)}}\,.\notag
\end{eqnarray}
This means that
the coefficients by
$\dhind 1$ and $\dhind 2$ in the above expression on the left
and on the right are equal.
Clearly
$\dst_3^-K\,w_\ka= (p+2)\dhind{1}\dst_3^-f_2 w_\ka -
\dhind{2}\dst_3^-(f_{12}  - f_1 f_2) w_\ka$.
Thus the equality of coefficients by $\dhind 1$ gives
\begin{equation}
  \notag
  h_1(h'+2)\dst_3^-f_2 w_\ka -n(p+2)\dst_3^-f_2
  w_\ka
= h_1(h'+2)e_0^- w_{(n,0,0)}\,.
\end{equation}
As $\wtt \dst_3^-f_2 w_\ka = (p+1,-2)$ we come to the equation
\begin{equation}
  \label{eq:d1}
  (p+1-n)\,\dst_3^-f_2 w_\ka = (p+1)\,e_0^- w_{(n,0,0)}\,.
\end{equation}
For  $\dhind 2$ taking into account weights we get
\begin{equation}\notag
\left(f_1 (p+2)\dst_3^-f_2
+n\dst_3^-B
+n\dst_3^-(f_{12}  - f_1 f_2)
\right)w_\ka
=
f_1(p+2)e_0^- w_{(n,0,0)}
+e_0^- w_{(n-1,1,0)}
\end{equation}
Or $\s\dst_3^-(n\,B+n(f_{12}-f_1f_2)+(p+2)f_1f_2)w_\ka=e_0^- w_{(n-1,1,0)}+
(p+2)f_1e_0^- w_{(n,0,0)}$, which is the same as
\begin{equation}
  \label{eq:d2}
(p+2)\,\dst_3^-\left(nf_{12}+f_1f_2
\right)w_\ka
=
e_0^- w_{(n-1,1,0)}+
(p+2)f_1e_0^- w_{(n,0,0)}\-.
\end{equation}

Notice that
\begin{equation}
  \label{eq:4.14}
\dst_3^-f_2 w_\ka=\dst_{31}^-f_{12}(\dst_1^+x_1^p)\,
\s\text{ and }\s
\dst_3^-f_{12}w_\ka=(\dst_{31}^-f_1-(p+1)\dst_{32}^-
)f_{12}(\dst_1^+x_1^p)
\,,
\end{equation}
therefore from \eqref{eq:d1} it follows that
\begin{equation}
  \label{eq:4.13'}
e_0^- w_{(n,0,0)}=\frac{p+1-n}{p+1}\,\dst_{31}^-f_{12}(\dst_1^+x_1^p)\,.
\end{equation}
On the other hand
it is clear that $\wtt w_{(n,0,0)}=(p+1,0)$ and
$w_{(n,0,0)}\in\La^-_2\La^+_2 V$,
hence by Proposition~\ref{P:hiv-p} we get
$w_{(n,0,0)}= c_1\,\dst^-_1\cycl{\dst^-}{\dst^+}{\dst^+}x_1^{p}$.
Then a straightforward calculation shows that
\[
e_0^- w_{(n,0,0)}=c_1\,\dst_{13}^-f_{12}(\dst_1^+x_1^p)
=-c_1\,\dst_{31}^-f_{12}(\dst_1^+x_1^p)\,.
\]
Comparing with \eqref{eq:4.13'}
we get $c_1=-\tfrac{p+1-n}{p+1}$.

Similarly from \eqref{eq:d2} we get
\begin{equation}
  \label{eq:d2'}
e_0^- w_{(n-1,1,0)}=
(p+2)\left(\dst_{31}^-(n+1+c_1)f_1+
\dst_{32}^-(1-n(p+1)+c_1)
\right)f_{12}(\dst_1^+x_1^p)\,.
\end{equation}
Also from Proposition~\ref{P:hiv-p} we get
\begin{equation}
  \label{eq:-1}
w_{\ka-(1)}=w_{(n-1,1,0)}=
      c_2\, \dst^-_{12}\cycl{\dst^+}{\dst^+}{x}x_1^{p-1}
+
      c_3\, \cycl{\dst^-}{\dst^-}{\dst^+_{12}\,x}x_1^{p-1}
\,.
\end{equation}
Hence $ e_0^- w_{(n-1,1,0)}$ can be written as follows.
\begin{equation}
  \label{eq:4.15'}
  \begin{array}[3]{rcl}
 e_0^- w_{(n-1,1,0)}=&\,\,\,&
 c_2\,\dst^-_{12}((p-1)\dst_1^+x_3^{2}x_1^{p-2}-p\,\dst_3^+x_3x_1^{p-1})\\
          &+&
 c_3\,\dst^-_{12}((p-1)\dst_1^+x_3^{2}x_1^{p-2}+\dst_3^+x_3x_1^{p-1})\\
     &    + &
 c_3\,\dst^-_{23}(\s\s\,\,p\,\dst_1^+x_3x_1^{p-1}+\dst_3^+x_1^{p})\\
     &    + &
 c_3\,\dst^-_{31}((p-1)\dst_1^+x_2x_3x_1^{p-2}+\dst_2^+x_3x_1^{p-1}+
                                               \dst_3^+x_2x_1^{p-1})\,.
\end{array}
\end{equation}
\/From this and \eqref{eq:d2'} we conclude that
\begin{eqnarray*}
  (p-1)(c_2+c_3)&=&0\,,\\
   -\, p\,c_2 + c_3&=&0\,.
\end{eqnarray*}
We see that either $w_{(n-1,1,0)}=0$ or $p=1$, $n=1$,
because $n=\ka_1\leq p$.

Let us consider the first alternative.
Here \eqref{eq:d2'} reduces to the equation
\begin{eqnarray*}
(n+1+c_1)\,\dst_{31}^-f_1f_{12}(\dst_1^+x_1^p)
&=&(n(p+1)-1-c_1)\,
\dst_{32}^-
f_{12}(\dst_1^+x_1^p)\,,
\end{eqnarray*}
which implies $n+1+c_1=n(p+1)-1-c_1=0$, therefore $n(p+2)=0$, hence
$n=0$ and $\ka=(0,1,0)$. This clearly gives us Theorem~\ref{T:qs-p}(4).

For the second alternative $c_1=-\tfrac{1}{2}$, and we get
$c_2=c_3=\tfrac{9}{2}$ matching \eqref{eq:d2'} with \eqref{eq:4.15'}.
Let us observe that $\wtt w=\wtt w_\ka+(0,-1)=(1,0)$ and
\[
w=D_1D_2w_\ka + D_1 w_{(1,0,0)} +D_2w_{(0,1,0)} +w_{(0,0,0)}\,,
\]
with no other terms possible. For $w_{(0,0,0)}$ we have
$w_{(0,0,0)}\in \La_3^-\La_3^+V$, hence
$w_{(0,0,0)}=c_4\dst_{123}^-\dst_{123}^+x_1$.
Therefore we have the information to compute $e^-_0w$ quite explicitly.
It takes some effort to check that it is not equal to zero whatever
$c_4$. We get no quasi-singular vectors.

{\bf Subcase}: $\ka_2=0$.\\
Now $\ka=(N,0,0)$, therefore
\begin{eqnarray*}
N\,D_1^{N-1}\{2\}
\dst_3^-w_\ka
&\equiv&  \sum_{|\al|=N-1}D\{2\}^\al e_0^- w_\al
\bmod  \sdeg(< N-1)\,.
\end{eqnarray*}
The expression \eqref{eq:-1} still holds for $w_{\ka-(1)}$,
and similarly we cannot have $\dst_{12}^-$ in $e^-_0w_{\ka-(1)}$,
therefore we are bound to conclude that $p=1$, $N=1$.
This leads us to the quasi-singular  vector from Theorem~\ref{T:qs-p}(7).\\

{\bf Case} 2: $\underline{w_\ka=\dst_1^-\dst_1^+\,x_1^p}$.
\smallskip\\
Here $\ka_1\leq p+2$, $\ka_2=0$ by Remark~\ref{R:4.3}.
\bLe
$\ka_3=0$. For any $\ka_1\leq p+2$ the vector exists.
\eLe
\begin{proof}
It is clear from \eqref{eq:4.4} that
if $\ka_3\neq 0$, then
$e_0^-w_{\ka{\scriptstyle{-(3)\,}}}\neq 0$ and in this case
$\wtt w_{\ka{\scriptstyle{-(3)\,}}}= (p+2,1,0)$.
A vector with such weight is not present in the list of
Proposition~\ref{P:hiv-q}. Hence $\ka_3=0$.
Now the expression given in Theorem~\ref{T:qs-p}(1) shows that for
any $\ka=(\ka_1,0,0)$ a vector $w$ exists.
\end{proof}

{\bf Case} 3:
$\underline{w_\ka=\dst_{123}^+,\,\,\dst_{1}^-\dst_{123}^+,\,\,
\dst_{12}^-\dst_{123}^+,\,\,\dst_{123}^-\dst_{123}^+}$.\smallskip\\
Form \eqref{eq:4.1} we conclude that
$\ldeg w_\al \geq \ldeg w_\ka +2 \geq 5$
for any $\al \neq \ka$, and that $\wt w_\al = \wt w_\ka$.
One immediately checks that this leaves no options but $w_\al=0$.
Therefore $w=D^\ka w_\ka$.

For $w_\ka=\dst_{123}^-\dst_{123}^+$, it follows from Remark~\ref{R:4.3}
that  $\ka_1=\ka_2=0$. After that $\ka_3$ is arbitrary as the vectors
given by Theorem~\ref{T:qs-p}(2) show.

For the rest of the vectors $\dst_3^- w_\ka\neq 0$.
Now, because $w=w_{\text{\footnotesize{top}}}$,
from \eqref{eq:4.4} it follows that
$\ka_3=0$. Here $\ka_1,\ka_2\leq 1$ by Remark~\ref{R:4.3}, and
we are left with pretty limited choices. It is easy to check that
all what we get are vectors listed in Theorem~\ref{T:qs-p}(9) and (10).

We have gone through all the cases of Proposition~\ref{P:4.2}, thus
 the proof of Theorem~\ref{T:qs-p} is accomplished.

\section{Proof of Theorem~\ref{T:ss-p}.}
\label{sec:5}

We shall work with vectors given by Theorem~\ref{T:qs-p}.
Let us
tabulate the weights of the vectors belonging to various cases of the theorem:
\begin{align*}
\text{case}&\s\s\s p\s&\s\wtt\s\s&\s\s\s\!\!\wt\s&\wtg\s\,\\
(1)  &\s\s p\geq 0  & (2+p-n,0)\s\s  &\s\s\s \, 0 & -\tfrac{2(n+1)}{3}\,  \\
(2)  &\s\s p=0      & (0,n) \s\s   &\s\s\s \, 0 & -\tfrac{2(n+3)}{3} \,\\
(3)  &\s\s p=1      & (0,0)\s\s    &\s\s\s \, 1 & \,\,-\tfrac{5}{3} \s\s\,\\
(4)  &\s\s p\geq 0  & (p+1,0)\s\s    &\s\s\s \, 0 & \,\,-\tfrac{4}{3}\s\s \,\\
(5)  &\s\s p=1      & (0,0)\s\s    &\s\s\s \, 0 & \,\,-\tfrac{2}{3}\s\s \,\\
(6)  &\s\s p=1      & (1,0)\s\s    &\s\s\s \, 1 & \,\,-1 \s\s\,\\
(7)  &\s\s p=1      & (0,1)\s\s    &\s\s\s \, 0 & \,\,-\tfrac{4}{3} \s\s\,\\
(8)  &\s\s p=0      & (0,0)\s\s    &\s\s\s \, 1 & \,\,-1\s\s\, \\
(9)  &\s\s p=0      & (0,0)\s\s    &\s\s\s \, 2 & \,\,-2\s\s\, \\
(10) &\s\s p=0      & (1,0)\s\s    &\s\s\s \, 1 & \,\,-\tfrac{7}{3}\s\s\,
\end{align*}
There are  coincidences in the table, namely (1) for $n=1$ gives
the same values as (4) and that is all. We are to
check how
$e_0^+$ acts on a quasi-singular vector which has to be a weight
vector and a linear combination of the vectors given in the various
cases
of Theorem~\ref{T:qs-p}. We conclude that the only linear combination
to consider is (1) for $n=1$ and (4), the other cases are to be
considered separately. We will go case by case.\\

{\bf Case (1)}. \\
Here we consider the semi-singular vector of the form
%
\begin{eqnarray}
  \label{eq:5.1}
&& w_{(1)}=D_1^{n}u_1-a\,D_1^{n-1}u_2-b\,D_1^{n-2}u_3\,,\quad\text{ where }
\\
\notag&&
u_1=\dst_1^-\dst_1^+\,x_1^p\,,\qquad
u_2=\dst_1^-(\dst_\circ^-\dst_\circ^+)_{23}\dst_1^+\,x_1^p\,,  \qquad
u_3=\dst_{123}^-\dst_{123}^+\,x_1^p\,.\qquad\qquad
\end{eqnarray}
and $a=n(p+3)$,
$b=n^{[2]}(p+3)^{[2]}$.
By Proposition~\ref{P:eD} we can write
\begin{eqnarray}
  \label{eq:5.2}
  \hspace*{4ex} e_0^+\,w_{(1)} &=& D_1\{2\}^{n}e_0^+u_1
  -aD_1\{2\}^{n-1}(e_0^+u_2 +\tfrac{1}{p+3}\dst_3^+B\,u_1   )  \\
 && -bD_1\{2\}^{n-2}(e_0^+u_3 -\tfrac{1}{p+2}\dst_3^+B\,u_2   )  \notag
  +b(n-2)D_1\{2\}^{n-3}(\dst_3^+B\,u_3)\,.\notag
\end{eqnarray}
Let us compute the terms.
\bLe
\label{L:5.1}
{}
a. $e_0^+u_1=0$,\\
b. $e_0^+u_2 +\tfrac{1}{p+3}\dst_3^+B\,u_1 \equiv
-\dhind 2f_{12}(\dst_1^+\,x_1^p)
\,\,\bmod \sdeg (<1) $,\\
c. $e_0^+u_3 -\tfrac{1}{p+2}\dst_3^+B\,u_2 \equiv 0
\,\,\bmod \sdeg (\leq 1) $,\\
d. $\dst_3^+B\,u_3 \equiv 0 \,\,\bmod \sdeg (\leq 1) $.
\eLe
\begin{proof}
For a. we have
\begin{equation*}
 e_0^+\,u_1=e_0^+\dst_1^-\dst_1^+\,x_1^p=-f_2\,\dst_1^+\,x_1^p+0=0\,.
\end{equation*}
Now
\begin{equation*}
e_0^+\,u_2 =
e_0^+\dst_1^-(\dst_\circ^-\dst_\circ^+)_{23}\dst_1^+\,x_1^p
=-f_2(\dst_\circ^-\dst_\circ^+)_{23}\dst_1^+\,x_1-\dst_1^-f_{12}\dst_{31}^+\,x_1^p
=-p\,\dst_1^-\dst_{31}^+\,x_3x_1^{p-1}\,,
\end{equation*}
and
\begin{eqnarray*}
\tfrac{1}{p+3}\dst_3^+B\,u_1&=& \dst_3^+f_{12}(\dst_1^-\dst_1^+\,x_1^p)\\
&=&\dst_3^+\dst_1^-f_{12}(\dst_1^+\,x_1^p)+\dst_3^+\dst_3^-\dst_1^+\,x_1^p
\equiv
-\dhind 2f_{12}(\dst_1^+\,x_1^p)
\,\,\bmod \sdeg (<1)\,,
\end{eqnarray*}
hence b. follows. In calculation with c. and d. we shall
similarly move $\dst_3^+$ over $\dst_i^-$ that can make terms
with no more than one $\dhind k$, thus the statements follow.
\end{proof}

It follows from the lemma and \eqref{eq:5.2} that
\begin{equation}
  \label{eq:5.3}
e_0^+\,w_{(1)}  \equiv
 a D_1\{2\}^{n-1}\dhind 2 f_{12}(\dst_1^+\,x_1^p)
\,\,\bmod \sdeg (< n)\,.
\end{equation}
We can apply Proposition~\ref{P:Ds-a} with $\al=(n-1,0,0)$, $s=2$,
thus $b=k=0$ in the summation on the right hand side. Let us notice
that
\begin{eqnarray*}
& &\hspace*{-5em}f_1^i B^j
(h_1-i-j)^{[n{\scriptscriptstyle{-}}1{\scriptscriptstyle{-}}i{\scriptscriptstyle{-}}j]}
(h'\{2\}-j)^{[n{\scriptscriptstyle{-}}1{\scriptscriptstyle - }j]}
\dhind 2 f_{12}(\dst_1^+\,x_1^p)\notag\\
&=&
(p+1-i-j)^{[n{\scriptscriptstyle{-}}1{\scriptscriptstyle{-}}i{\scriptscriptstyle{-}}j]}
(p+2-j)^{[n{\scriptscriptstyle{-}}1{\scriptscriptstyle - }j]}
f_1^i B^j\dhind 2 f_{12}(\dst_1^+\,x_1^p) \\
&=&
(p+1-i-j)^{[n{\scriptscriptstyle{-}}1{\scriptscriptstyle{-}}i{\scriptscriptstyle{-}}j]}
(p+2-j)^{[n{\scriptscriptstyle{-}}1{\scriptscriptstyle - }j]}
\,
(p+2)^{[j]}
\,
f_1^i
\dhind 2
f_{12}^{j+1}(\dst_1^+\,x_1^p)\\
&=&
(p+1-i-j)^{[n{\scriptscriptstyle{-}}1{\scriptscriptstyle{-}}i{\scriptscriptstyle{-}}j]}
(p+2)^{[n{\scriptscriptstyle{-}}1
]}
(-i\dhind 1f_1^{i-1}+\dhind 2f_1^i)f_{12}^{j+1}(\dst_1^+\,x_1^p)
\,.
\end{eqnarray*}
We see that in calculating "$\bmod\, \dlxq{(n-2,1,1)}$" we need only
to consider the terms with $i,j=0,0$, $i,j=1,0$, $i,j=2,0$,
 and $i,j=1,1$, therefore
"$\bmod\, \dlxq{(n-2,1,1)}$"
\begin{eqnarray}
  \label{eq:5.4}
\hspace*{5em}e_0^+\,w_{(1)}  \equiv
C &&\hspace*{-1em}\left(\binom{n-1}{0}
(p+1)^{[n{\scriptscriptstyle{-}}1]}
\dhind 1^{n-1}(\,\dhind 2)\notag\right.\\
&&\hspace*{-1em}+\binom{n-1}{1}
\,\,p^{[n{\scriptscriptstyle{-}}2]}
\,\,\,\dhind 1^{n-2}\dhind 2(-\dhind 1+\dhind 2f_1 )\\
&&\notag
\hspace*{-1em}+\binom{n-1}{2}
(p-1)^{[n{\scriptscriptstyle{-}}3]}
\dhind 1^{n-3}\dhind 2^2(-2\dhind 1f_1+\dhind 2f_1^2)\\
&&\left.
\hspace*{-1em}+\binom{n-1}{1,1}
(p-1)^{[n{\scriptscriptstyle{-}}3]}
\dhind 1^{n-3}\dhind 2\dhind 3(-\dhind 1+\dhind 2f_1)f_{12}
\right)
f_{12}(\dst_1^+\,x_1^p)\notag\,,
\end{eqnarray}
where $C=a(p+2)^{[n-1]}\neq 0$ because $0<n\leq p+2$.
We should
keep in mind that some terms are absent for $n\leq 2$,
namely for $n=1$ all terms but the first disappear, and for $n=2$
we have just the first two terms left.

Clearly
for $n\geq 3$, thus $p\geq n-2\geq 1$,
we can rewrite \eqref{eq:5.4} as follows
\begin{eqnarray}
  \label{eq:5.5}
\hspace*{5em}e_0^+\,w_{(1)}  \equiv
A &&\hspace*{-1em}\left(
p^{[n{\scriptscriptstyle{-}}2]}(
(p+1)-(n-1))\dhind 1^{n-1}\dhind 2
\notag\right.\\
&&\hspace*{-1em}+(p-1)^{[n{\scriptscriptstyle{-}}3]}
(n-1)(p-(n-2))\dhind 1^{n-2}\dhind 2^2f_1\\
&&\left.
\hspace*{-1em}-(n-1)(n-2)
(p-1)^{[n{\scriptscriptstyle{-}}3]}
\dhind 1^{n-2}\dhind 2\dhind 3 f_{12}
\right)
f_{12}(\dst_1^+\,x_1^p)\notag\,.
\end{eqnarray}
Here
$p^{[n{\scriptscriptstyle{-}}2]}\neq 0$,
$(p-1)^{[n{\scriptscriptstyle{-}}3]}\neq 0$,
and
we notice that $f_{12}^2(\dst_1^+\,x_1^p)\neq 0$ because $p\geq 1$.
We conclude that $e_0^+\,w =0$ is not possible when $n\geq 3$.

It is easy to see from \eqref{eq:5.4} that $n=1$ is impossible too.
We are left with $p=0$, $n=2$, which is in fact the case (ii) of
Theorem~\ref{T:ss-p}, we should only notice that the vector
$\dst^-_{123}\dst^+_{123}$ is itself semi-singular, it
could be either added or not in the expression.\\

{\bf Cases (1)\&(4)}. \\
We are to consider a linear combination $w=A w_{(1)} + B w_{(4)}$
where $w_{(1)}$ is given by \eqref{eq:5.1} with $n=1$ and $w_{(4)}$
is the vector of the case (4) of Theorem~\ref{T:qs-p}, thus
\begin{align}
  \label{eq:5.6}
w_{(4)}&=D_2\,v_1 - v_1\,,\,
\text{ where } \\ \notag
v_1&=(\dst^-_\circ \dst^+_\circ )_{12}\,x_1^{p}
        +p(\dst^-_\circ \dst^+_{1}x_\circ )_{12}\,x_1^{p-1},\,\\
v_2&=\dst^-_1(\dst^-_\circ\dst^+_\circ)_{23}\,{\dst^+_1}x_1^{p}.\notag
\end{align}
One can check that
$e_0^+v_1=-(p+2)f_{12}\dst_1^+\,x_1^p$,
$e_0^+v_2=-p \dst_1^-\dst_{31}^+\,x_3x_1^{p-1}$ and
\begin{eqnarray}
  \label{eq:5.7}
e_0^+w_{(4)}&=&-(p+2)\dhind 2f_{12}\dst_1^+\,x_1^p-
\dst_{3}^+f_2v_1-e_0^+v_2 \notag\\
&=&-(p+2)\dhind 2f_{12}\dst_1^+\,x_1^p-\dst_{3}^+
((\dst^-_\circ \dst^+_\circ )_{13}\,x_1^{p}
        +p(\dst^-_\circ \dst^+_{1}x_\circ )_{13}\,x_1^{p-1})-e_0^+v_2 \\
&=&-(p+1)\dhind 2f_{12}\dst_1^+\,x_1^p
-(p+1)\dst_3^-\dst_{31}^+\,x_1^{p}
+p\,\dst_1^-\dst_{31}^+\,x_3x_1^{p-1}\notag\,.
\end{eqnarray}
On the other hand as we already calculated above
\begin{eqnarray}
  \label{eq:5.8}
e_0^+w_{(1)}&=&-\dst_{3}^+B\,u_1-(p+3)u_2 \notag\\
  &=&(p+3)\dhind 2f_{12}\dst_1^+\,x_1^p
+(p+3)\dst_3^-\dst_{31}^+\,x_1^{p}
+2(p+3)p\,\dst_1^-\dst_{31}^+\,x_3x_1^{p-1}\,.\qquad
\end{eqnarray}
Therefore $e_0^+w=0$ implies that $A(p+3)=B(p+1)$ and $A\,2(p+3)p=B\,p$.
We conclude that $p=0$, $B=3A$, and the vector is
proportional to the one of the case (iii) of Theorem~\ref{T:ss-p}.\\

In the remaining cases (2), (3) and (5)-(10) the calculations are
straightforward, in (2), (3) and (5)-(7) we get no semi-singular
vectors, the case (8) corresponds to (i), (9) to (iv), and (10) to (v).

\section{Proof of Theorem~\ref{T:qs-q}.}
\label{sec:6}

As before, our considerations are based on
\eqref{eq:w-gen}-\eqref{eq:w-ntp},
Proposition~\ref{P:e-tp} and Remarks~\ref{R:w-tpcf}, \ref{R:4.1}, \ref{R:4.3}
We follow the same strategy that was used for finding quasi-singular
vectors in Section~\ref{sec:4}, and
we keep the notations and conventions made at the beginning of
Section~\ref{sec:4}. But we are now in the context of
Theorem~\ref{T:qs-q}, in particular $q\geq 1$.

Let us remind that by Proposition~\ref{P:e-tp}
the possible top level coefficients are listed in
Corollary~\ref{C:qs-deg0-q}. Remark~\ref{R:4.1}
helps us to describe them quite explicitely.
\bPr\label{P:6.1}
{\rm (1):}
If
$\,\,w_\ka = \pt_3^q\,, \,\, \Dep(\pt_*)\pt_3^{q-1}\,,  $
or
$\mx{ $
\Dem(\Dep(\pt_*)\pt_*)\pt_3^{q-2}\,, \s q\geq 2
\,$},$
then
$\,I_{\text{top}}=\{ \ka \}$.\smallskip
\newline
{\rm (2):}
If
$\,\,w_\ka =
\dst^-_1\Dep(\pt_*)\pt_3^{q-1}
\,$
then either
$\,I_{\text{top}}=\{ \ka \}$ or
$\,I_{\text{top}}=\{ \ka, \si \}$,
where \\
$\mx{$\si=\ka +(-1,0,1)$}$
and $\,w_\si=c_\si\Dem(\Dep(\pt_*)\pt_*)\pt_3^{q-2}\,,
   \,c_\si\in \CC , \s q\geq 2$.\smallskip
\newline
{\rm (3):}
If $q=1$ and
$w_\ka = \dst^+_1\Dep(\pt_*)\,,  \s
 (\dst^-_\circ \dst^+_\circ )_{12}\Dep(\pt_*)\,, \s
    \cycl{\dst^-}{\dst^-}{\dst^+}\Dep(\pt_*)\,,
$
then $\mx{$I_{\text{top}}=\{ \ka \}$}$.\smallskip
\newline
{\rm (4):}
If $q=1$ and $w_\ka =\dst^-_{12}\dst^+_1\Dep(\pt_*)
$,
then either
$\,I_{\text{top}}=\{ \ka \}$ or
$\,I_{\text{top}}=\{ \ka, \si \}$,
where $\,\si=\ka +(-1,0,1)$
and $\,w_\si=\cycl{\dst^-}{\dst^-}{\dst^+}\Dep(\pt_*)
$.
\smallskip
\newline
{\rm (5):}
If $q=1$ and $w_\ka =\dst^-_1\dst^+_1\Dep(\pt_*)$,
then either
$\,I_{\text{top}}=\{ \ka \}$ or
$\,I_{\text{top}}=\{ \ka, \si \}$,
where $\,\si=\ka +(-1,1,0)$
and $\,w_\si=(\dst^-_\circ \dst^+_\circ )_{12}\Dep(\pt_*)$.
\ePr
\begin{proof}
What we need is to look at $s\ell (3)$-weights of vectors listed in
Corollary~\ref{C:qs-deg0-q} and use Remark~\ref{R:4.3}. We leave the
details to the reader.
\end{proof}
\bLe\label{L:6.1} In the conditions of Proposition~\ref{P:6.1}:\\
1. Whatever $I_{\text{top}}$ and $\ka$, one has
\begin{equation}
  \label{eq:4.4-}
\sum_{|\al|=N-1}D\{2\}^\al e_0^- w_\al
\equiv
\ka_3\,D^{\ka-(3)}\{1\}\,\dst_3^- w_{\,\ka{\scriptstyle{-(3)\,}}}
\s \bmod \,\lxl{\,\ka{\scriptstyle{-(3)\,}}}\,,
\end{equation}
2.  Whatever $I_{\text{top}}$, if  $\ka_3=0$ then
\begin{equation}
 \label{eq:6.1}
  \begin{array}[r]{r}
\sum_{|\al|=N-1}D\{2\}^\al e_0^- w_\al
\equiv \hspace{22em}\\
\ka_2 \hat{\partial}^{\ka-(2)}\dst_3^-f_2
((h_1+1)^{[\ka_1]}-\ka_1h_1^{[\ka_1-1]})
  {h'}^{[\ka_1]} (h_2-1)^{[\ka_2-1]}
w_\ka
\, \bmod \lxl{\ka{\scriptstyle{-(2)}}},
  \end{array}\end{equation}
\eLe
\begin{proof}
Let us notice that the
general formula \eqref{eq:4.2}
is valid in our context. To compute the RHS we
use  Proposition~\ref{P:eD} for $e_0^-D^\ka w_\ka $ and
$e_0^-D^\si w_\si $.

But clearly
$\si{\scriptstyle{-(3)\,}}<_{\mx{\small{lex}}}{\,\ka{\scriptstyle{-(3)\,}}}$
so we get (1). For (2) we have to check that
$\si{\scriptstyle{-(3)\,}}$ and
$\mx{$\si{\scriptstyle{-(2)\,}}$}$ are
strictly less than $\ka{\scriptstyle{-(2)\,}}$,
which is clear.
Then we use Corollaries~\ref{C:3.6},~\ref{C:Klex}.
\end{proof}

Now we begin to consider  various possibilities
for $w_\ka
$ { \bf case by case}
according to Proposition~\ref{P:6.1}.
\\

{\bf Case} 1(i): $\underline{w_\ka=\pt_3^q\,}$.\\
By Remark~\ref{R:4.3} $\ka_1=0$.
\bLe\label{L:6.2}
$\ka_3=0$.
\eLe
\begin{proof}
From
\eqref{eq:4.4-} it follows that
either
$\ka_3=0$ or $e_0^-w_{\ka{\scriptstyle{-(3)\,}}}\neq 0$.
Checking with the list of Proposition~\ref{P:hiv-q} with the weights
in mind we see that the latter is impossible.
\end{proof}
We conclude that $\ka=(0,N,0)$ and now \eqref{eq:6.1} is valid with
the RHS equal to
\[
\, \hat{\partial}^{\ka-(2)}\dst_3^-f_2(h_2-1)^{[\ka_2-1]}
w_\ka=
- q^{[N]}\, \hat{\partial}^{\ka-(2)}\dst_3^-\pt_2\pt_3^{q-1}\,,
\]
and $q^{[N]}\neq 0$ because $ N \leq q$.

The weight of $w_{\,\ka{\scriptstyle{-(2)\,}}}$ is easy to determine and
Proposition~\ref{P:hiv-q} provides a vector of this weight
\begin{equation}
  \label{eq:6.2}
w_{\,\ka{\scriptstyle{-(2)\,}}}=
c_1\dst^-_1\Dep(\pt_*)\pt_3^{q-1}+
               c_2\Dem(\dst^+_1\pt_*)\pt_3^{q-1}\,.
\end{equation}
We can
assume that $c_1=0$ because $e_0^-\dst^-_1\Dep(\pt_*)\pt_3^{q-1}=0$.
Then $e_0^-w_{\,\ka{\scriptstyle{-(2)\,}}}=
c_2((q-1)\Dem(\pt_*)\pt_2\pt_3^{q-2}+\dst^-_3\pt_2\pt_3^{q-1})$.
Substituting this into \eqref{eq:6.1} we get
$q=1$, $N=1$, and $c_2=-1$.
We have come to the vector from
case (5) of Theorem~\ref{T:qs-q}.\\

{\bf Case} 1(ii): $\underline{w_\ka=\Dep(\pt_*)\pt_3^{q-1}\,}$.\\
As $\wtt w_\ka=(0,q-1)$,  Remark~\ref{R:4.3} implies that $\ka_1=0$.
Let us prove that $\ka_3=0$.

We shall use \eqref{eq:4.4-}. Let us consider what is possible for
$w_{\,\ka{\scriptstyle{-(3)\,}}}$.
By Proposition~\ref{P:hiv-q} and with weights in mind we get
\begin{equation}
  \label{eq:6.3}
w_{\,\ka{\scriptstyle{-(3)\,}}}=
c_1(\dst^-_\circ \dst^+_\circ )_{12}\Dep(\pt_*)\pt_3^{q-1}
+c_2\cycl{\dst^-}{\dst^+}{\dst^+} \pt_3^q\,.
\end{equation}
Immediate calculation shows that
\begin{equation}
  \label{eq:6.3'}
\begin{array}{crl}
e_0^-(\dst^-_\circ \dst^+_\circ )_{12}\Dep(\pt_*)\pt_3^{q-1}&=&
-(q-1)(\dst^-_1\Dep(\pt_*)\pt_1
+\dst^-_2\Dep(\pt_*)\pt_2)\pt_3^{q-2}\,,\\
e_0^-\cycl{\dst^-}{\dst^+}{\dst^+} \pt_3^q &=&
-q\Dem(\dst^+_3\pt_*)\pt_3^{q-1}   + \dst^-_3(\ldots)\,.
\end{array}
\end{equation}
It follows from  \eqref{eq:4.4-} that the only possibility is
$\ka_3=c_1=c_2=0$.

We conclude that $\ka=(0,N,0)$. Here
$1\leq N \leq q-1$ by Remark~\ref{R:4.3}, in particular $q\geq 2$.
Again we can use
\eqref{eq:6.1}, where the coefficient by $\hat{\partial}^{\ka-(2)}$
in the RHS is
\begin{equation}
  \label{eq:6.4}
\dst^-_3f_2(h_2-1)^{[\ka_2-1]} w_\ka=
-(q-1)^{[N]}\dst^-_3\Dep(\pt_*)\pt_2\pt_3^{q-1}\,,
\end{equation}
and of course $(q-1)^{[N]}\neq 0$.
On the other hand from Proposition~\ref{P:hiv-q} it follows that
\begin{equation}
  \label{eq:6.5}
w_{\,\ka{\scriptstyle{-(2)\,}}}=c\Dem(\dst^+_1\Dep(\pt_*)\pt_*)\pt_3^{q-2}\,,
\end{equation}
and that implies
\begin{equation}
  \label{eq:6.5'}
e_0^-w_{\,\ka{\scriptstyle{-(2)\,}}}
=-c(q-2)\Dem(\dst^+_1\Dep(\pt_*)\pt_*)\pt_2\pt_3^{q-3}+\dst^-_3(\ldots)\,.
\end{equation}
Substituting results of \eqref{eq:6.4} and \eqref{eq:6.5}
into \eqref{eq:6.1}
we arrive at a contradiction. Therefore no $w$
exists in this case. \\

{\bf Case}  1(iii): $\underline{w_\ka=\Dem(\Dep(\pt_*)\pt_*)\pt_3^{q-2}\,}$.\\
Now $\wtt w_\ka=(0,q-2)$ and again
Remark~\ref{R:4.3} implies that $\ka_1=0$ and
that for $q=2$ also $\ka_2=0$.
Let us prove that $\ka_2=0$ for $q>2$ as well.
Then we come to the vector given by Theorem~\ref{T:qs-q}(1).

Denote $\ka=(0,s,t)$ and suppose $s\geq 1$.
\/From Proposition~\ref{P:hiv-q} with our weigths in mind we get
\begin{equation}
  \label{eq:6.6'}
w_{\,\ka{\scriptstyle{-(3)\,}}}=
c\,\cycl{\dst^-}{\dst^-}{\dst^+}\Dep(\pt_*)\pt_3^{q-1} \,,
\end{equation}
and $w_{\,\ka{\scriptstyle{-(2)\,}}}=0$.
By Proposition~\ref{P:Ds-a}
\begin{equation*}
D^\ka w_\ka=
D_2^s D_3^t w_\ka=
\sum_{k=0}^{s}
\binom{s}{k}\hat{\partial}_2^{s-k}
\hat{\partial}_3^{t+k}
(q-2-k)^{[s-k]}f_2^k w_\ka\,,
\end{equation*}
therefore
\begin{equation}
  \label{eq:6.7'}
e_0^-D^\ka w_\ka\,=\,
\sum_{k=0}^{s}
\binom{s}{k}(t+k)
\hat{\partial}_2^{s-k}
\hat{\partial}_3^{t+k-1}
(q-2-k)^{[s-k]}(-\dst^-_3 f_2^k  w_\ka )\,.
\end{equation}
Similarly
\begin{eqnarray}
  \label{eq:6.8'}
e_0^-D^{\ka-(3)}
 w_{\ka{\scriptstyle{-(3)}}}=
c\sum_{k=0}^{s}
\binom{s}{k}
\hat{\partial}_2^{s-k}
\hat{\partial}_3^{t+k-1}
(q-1-k)^{[s-k]}
f_2^k e_0^- w_{\ka{\scriptstyle{-(3)}}} \\
\qquad
+c\sum_{k=0}^{s}
\binom{s}{k}(t+k-1)
\hat{\partial}_2^{s-k}
\hat{\partial}_3^{t+k-2}
(q-1-k)^{[s-k]}
(-\dst^-_3 f_2^k  w_{\ka{\scriptstyle{-(3)}}})\,.
\notag
\end{eqnarray}
Because $\ka_1=0$ and
$w_{\,\ka{\scriptstyle{-(2)\,}}}=0$
we have
\begin{equation}
  \label{eq:6.9'}
e_0^-D^\ka w_\ka+
e_0^-D^{\ka-(3)}w_{\ka{\scriptstyle{-(3)}}}
\equiv 0 \quad\bmod \sdeg (<N-1)\,.
\end{equation}
Looking at coefficients by $\hat{\partial}_2^{s-k}
\hat{\partial}_3^{t+k-1}$ for $k=0,1$ in \eqref{eq:6.7'}, \eqref{eq:6.8'}
we get from  \eqref{eq:6.9'}
\begin{eqnarray*}
(q-1)^{[s]}\,e_0^- w_{\ka{\scriptstyle{-(3)}}}
&=& t\,(q-2)^{[s]}\,\dst^-_3   w_\ka \,,    \\
(q-2)^{[s-1]}f_2\,e_0^- w_{\ka{\scriptstyle{-(3)}}}
&=&(t+1)(q-3)^{[s-1]}\dst^-_3 f_2  w_\ka \,.
\end{eqnarray*}
But $e_0^- w_{\ka{\scriptstyle{-(3)}}}=(q-1)\dst^-_3
\Dem(\Dep(\pt_*)\pt_*)\pt_3^{q-2}$, therefore we come to
\begin{eqnarray*}
c\,(q-1)^{2}
&=&t\,(q-1-s)\,,    \\
c\,(q-1)(q-2)
&=&(t+1)(q-1-s) \,.
\end{eqnarray*}
Clearly the system implies $(q-2)\,t=(q-1)(t+1)$ which has no positive
solutions. Thus $\ka_2 > 0$ is impossible, as we claimed.\\

{\bf Case} 2: $\underline{w_\ka=\dst^-_1\Dep(\pt_*)\pt_3^{q-1}\,}$.\\
We have $\wtt w_\ka=(1,q-1)$ and Remark~\ref{R:4.3} implies that
$\ka_1=0 \text{ or } 1$, and $\ka_2 \leq q-1$.

Let us  prove that $\ka_3=0$.
Equation \eqref{eq:4.4-} is valid here with
$\dst^-_3 w_\ka=\dst^-_{31}\Dep(\pt_*)\pt_3^{q-1}$ in the RHS.
Thus for $w_{\,\ka{\scriptstyle{-(3)\,}}}$ we get
\begin{equation*}
w_{\,\ka{\scriptstyle{-(3)\,}}}
=-c_1\dst^-_{12} \dst^+_{1}\Dep(\pt_*)\pt_3^{q-1}
+\,c_2\,\dst^-_1\cycl{\dst^-}{\dst^+}{\dst^+} \pt_3^q\,,
\end{equation*}
which coincides with the RHS of \eqref{eq:6.3} multiplied by
$\dst^-_1$. Therefore calculating $e_0^-w_{\,\ka{\scriptstyle{-(3)\,}}}$
we can use the results in the RHS of \eqref{eq:6.3'} multiplied by $\dst^-_1$,
and that gives
\begin{equation*}
  \begin{array}{crlll}
e_0^-w_{\,\ka{\scriptstyle{-(3)\,}}}&=&
c_1(q-1)\,
\dst^-_{12}\Dep(\pt_*)\pt_2\pt_3^{q-2}
-c_2\, q\,\dst^-_1 \Dem(\dst^+_3\pt_*)\pt_3^{q-1}   &+ &\dst^-_{13}(\ldots)\,\\
&= &
c_1(q-1)\,
\dst^-_{12}\Dep(\pt_*)\pt_2\pt_3^{q-2}
-c_2 \,q\,\dst^-_{12}\dst^+_3\pt_2\pt_3^{q-1}  & + &\dst^-_{13}(\ldots)\,.
\end{array}
\end{equation*}
We conclude that $q=1$, $c_2=0$, but then $e_0^-w_{\,\ka{\scriptstyle{-(3)\,}}}=0$
hence $\ka_3=0$.

We can now use \eqref{eq:6.1} where in the RHS we have
\begin{equation}
  \label{eq:6.6}
\dst_3^-f_2\, w_\ka=(q-1)\dst^-_{13}\Dep(\pt_*)\pt_2\pt_3^{q-2}\,,
\end{equation}
and for $w_{\,\ka{\scriptstyle{-(2)\,}}}$
we get
\begin{equation}
  \label{eq:6.7}
w_{\,\ka{\scriptstyle{-(2)\,}}}=c\,\dst^-_1\Dem(\dst^+_1\Dep(\pt_*)\pt_*)\pt_3^{q-2}\,,
\end{equation}
which of course is equal to the RHS
of  \eqref{eq:6.5} multiplied by $\dst^-_1$.

\bLe\label{L:6.3}
If $q > 2$ then $\ka_2=0$.
\eLe
\begin{proof}
To calculate $e_0^-w_{\,\ka{\scriptstyle{-(2)\,}}}$ we can use
\eqref{eq:6.5'}, and we get
\begin{equation}
  \label{eq:6.8}
e_0^-w_{\,\ka{\scriptstyle{-(2)\,}}}
=-\,c\,(q-2)\dst^-_1\Dem(\dst^+_1\Dep(\pt_*)\pt_*)\pt_2\pt_3^{q-3}
+\dst^-_{13}(\ldots)\,.
\end{equation}
Comparing it with \eqref{eq:6.6} we see that if $q> 2$ then $c=\ka_2=0$.
\end{proof}

Notice that if $\ka_2=0$ then $\ka_1=1$ because $|\ka\,|\geq 0$,
and for $\ka=(1,0,0)$,  $q\geq 1$ the vector is given
by  Theorem~\ref{T:qs-q}(2). Because of Lemma~\ref{L:6.3},
we are left to consider the situation where $q=2$,  $\ka_2= 1$,
 $\ka_1 \leq 1$.
Here
for $\ka=(0,1,0)$ the vector is given by  Theorem~\ref{T:qs-q}(4)
and for $\ka=(1,1,0)$  by  Theorem~\ref{T:qs-q}(3).\\

{\bf Case}   3(i): $\underline{w_\ka=\dst^+_1\Dep(\pt_*)\,}$.\\
Here $\ka_2=0$, $\ka_1\leq 1$ by Remark~\ref{R:4.3}. From
Proposition~\ref{P:hiv-q} we get
$w_{\,\ka{\scriptstyle{-(3)\,}}}=c\,\dst^-_{1}\dst^+_{123}\pt_3$,
then \eqref{eq:4.4-} shows that $c=\ka_3=0$.

We are bound to have  $\ka=(1,0,0)$, and then
the vector is given by  Theorem~\ref{T:qs-q}(6).\\

{\bf Case}  3(ii):
 $\underline{w_\ka=(\dst^-_\circ \dst^+_\circ )_{12}\Dep(\pt_*)\,}$.\\
We have $\wtt w_\ka=(0,1)$, thus  $\ka_1=0$, $\ka_2\leq 1$ by Remark~\ref{R:4.3}.
Clearly  $\wtt w_{\,\ka{\scriptstyle{-(3)\,}}}=(0,2)$,
$\ldeg w_{\,\ka{\scriptstyle{-(3)\,}}}=5$,
$\wt w_{\,\ka{\scriptstyle{-(3)\,}}}=1$,
therefore from Proposition~\ref{P:hiv-q} we get
$w_{\,\ka{\scriptstyle{-(3)\,}}}=c\,\dst^-_{12}\dst^+_{123}\pt_3$.
Now \eqref{eq:4.4-} shows that $c=\ka_3=0$.
We come to  Theorem~\ref{T:qs-q}(7).\\

{\bf Case} 3(iii):
$\underline{w_\ka=\cycl{\dst^-}{\dst^-}{\dst^+}\Dep(\pt_*)\,}$.\\
Clearly $\ka_1=0$, $\ka_2=0$ here, and we come to vectors given by
Theorem~\ref{T:qs-q}(8).\\

{\bf Case}  4: $\underline{w_\ka=\dst^-_{12}\dst^+_1\Dep(\pt_*)\,}$.\\
Here  $\ka_1 \leq 1$, $\ka_2\leq 1$. From Proposition~\ref{P:hiv-q} we get
$w_{\,\ka{\scriptstyle{-(3)\,}}}=0$ because there is no vector with the
appropiate weight and $\ldeg w_{\,\ka{\scriptstyle{-(3)\,}}}=6$,
hence  $\ka_3=0$. Similarly $w_{\,\ka{\scriptstyle{-(2)\,}}}=0$
and $\ka_2=0$. Then $\ka=(1,0,0)$ and we have the vector
in Theorem~\ref{T:qs-q}(9).\\

{\bf Case}  5: $\underline{w_\ka=\dst^-_1\dst^+_1\Dep(\pt_*)\,}$.\\
We see that $\ka_2 =0$, $\ka_1\leq 2$ and $\ka_3=0$ because
no $w_{\,\ka{\scriptstyle{-(3)\,}}}$ could be found.
We have come to  Theorem~\ref{T:qs-q}(10).\\

This finishes the proof of Theorem~\ref{T:qs-q}.

\section{Proof of Theorem~\ref{T:ss-q}.}
\label{sec:7}

We shall denote $w_{(i)}$ the vector in the form given by the case (i)
of Theorem~\ref{T:qs-q}. Considering the weights of these vectors we
see that equal weights are relatively rare, therefore few linear combinations
are to be considered. Namely these linear combinations are:
 \alphaparenlist
\begin{enumerate}
\item $w_{(1)}|_{ n=1}\,$ and $w_{(2)}$,
\item $w_{(2)}|_{ q=1}\,$ and $w_{(8)}|_{n=0}\,$,
\item $w_{(7)}\,$ and $w_{(10)}|_{ n=1}\,$,
\item $w_{(8)}|_{  n=1}\,$ and $w_{(9)}$.
 \end{enumerate}
According to the definition we shall consider the action of $e^+_0$ on
$w_{(i)}$ going through cases $i=1,\ldots, 10$ Theorem~\ref{T:qs-q}
with the special regard to possible linear combinations.\\

{\bf Case}  (1):
According to the formula of Proposition~\ref{P:eD}
\begin{eqnarray}
  \label{eq:7.1}
e_0^+w_{(1)}&=&\s D_3^n\, e_0^+ \Dem(\Dep(\pt_*)\pt_*)\pt_3^{q-2} - \notag\\
&&- n\,D_3^{n-1} \left(\dst^+_{3}\Dem(\Dep(\pt_*)\pt_*)\pt_3^{q-2} -\tfrac{1}{q-1}
 e_0^+ \cycl{\dst^-}{\dst^-}{\dst^+}\Dep(\pt_*)\pt_3^{q-1}\right) +\\
&&-n^{[2]}D_3^{n-2} \left(\tfrac{1}{q-1}
          \dst^+_{3}\cycl{\dst^-}{\dst^-}{\dst^+}\Dep(\pt_*)\pt_3^{q-1}
+\tfrac{1}{q^{[2]}} e_0^+ \dst^-_{123}\dst^+_{123}\pt_3^q\right)-\notag\\
&&-n^{[3]}
D_3^{n-3}\left(\tfrac{1}{q^{[2]}}\dst^+_{3}\dst^-_{123}\dst^+_{123}\pt_3^q\right)
\,. \notag
\end{eqnarray}
The term in the last line is clearly zero. Also
$e_0^+ \Dem(\Dep(\pt_*)\pt_*)\pt_3^{q-2}=0$.
Now along with calculating the action of $e_0^+$,
we shall move $\dst^+_{3}$ over products of $\dst^-_i$ at some
places. Namely
\begin{align}
\label{eq:7.2}
\dst^+_{3}\Dem(\Dep(\pt_*)\pt_*)\pt_3^{q-2}&= \\
(\dhind 1 \Dep(\pt_*)&\pt_2-\dhind 2 \Dep(\pt_*)\pt_1)\,
\pt_3^{q-2}-\Dem(\dst^+_{3}\Dep(\pt_*)\pt_*)\pt_3^{q-2}\,,\notag\\
e_0^+\cycl{\dst^-}{\dst^-}{\dst^+}\Dep(\pt_*)\pt_3^{q-1}&=\label{eq:7.3}\\
\dst^+_{3}\dst^+_{3}&\Dep(\pt_*)\pt_3^{q-1}+(q-1)
\Dem(\dst^+_{3}\Dep(\pt_*)\pt_*)\pt_3^{q-2}\,.\notag
\end{align}
Looking at the other terms in \eqref{eq:7.1} we conclude that
\begin{equation*}
e_0^+w_{(1)}=-n\,\dhind 3^{n-1}(\dhind 1 \Dep(\pt_*)\pt_2-\dhind 2 \Dep(\pt_*)\pt_1)
\pt_3^{q-2} + (\dots)\,,
\end{equation*}
there $(\dots)$ marks the terms with  $\sdeg$ less than $n$. This gives
$e_0^+w_{(1)}\neq 0$ for all possible $n\geq 0$, $q\geq 2$.\\

{\bf Case}  (2):
Similarly from Proposition~\ref{P:eD} it follows
\begin{equation}
  \label{eq:7.4}
e_0^+w_{(2)}=
D_1\{2\}\,e_0^+\dst^-_1\!\Dep(\pt_*\!)\pt_3^{q-1}
-\dst^+_{3}B \,\dst^-_1\!\Dep(\pt_*\!)\pt_3^{q-1}
+ e_0^+\cycl{\dst^-}{\dst^-}{\dst^+}\Dep(\pt_*\!)\pt_3^{q-1}
\,.
\end{equation}
We are to calculate the terms. For the first two terms we get
\begin{align}
  \label{eq:7.5}
D_1\{2\}\,e_0^+\dst^-_1\!\Dep(\pt_*\!)\pt_3^{q-1} =&
  \,D_1\{2\}\,(q-1)\Dep(\pt_*\!)\pt_2\pt_3^{q-2}=\\
&\,(q-1)(q+1)(\dhind 1 \Dep(\pt_*\!)\pt_2-\dhind 2 \Dep(\pt_*\!)\pt_1)\,
\pt_3^{q-2}\notag \,,\\
  \label{eq:7.6}
-\dst^+_{3}B \,\dst^-_1\!\Dep(\pt_*\!)\pt_3^{q-1}=&
\,(q-1)(\dhind 1 \Dep(\pt_*\!)\pt_2-\dhind 2 \Dep(\pt_*\!)\pt_1)\,\pt_3^{q-2}\\
&+(q+1)\dst^+_{3}\dst^+_{3}\Dep(\pt_*\!)\pt_3^{q-1}
 -(q-1)\Dem(\dst^+_{3}\Dep(\pt_*\!)\pt_*\!)\pt_3^{q-2}\,,\notag
\end{align}
and the last term has been calculated in \eqref{eq:7.3}.

\hspace{0em}From \eqref{eq:7.1}-\eqref{eq:7.6} we conclude that no linear
combinations of $w_{(2)}$ and $w_{(1)}|_{n=1}$ makes a semi-singular
vector, but
that
\[ w=
w_{(2)}|_{q=1} - 3 \cycl{\dst^-}{\dst^-}{\dst^+}\Dep(\pt_*\!)=
D_1\dst^-_1\!\Dep(\pt_*\!) - 2 \cycl{\dst^-}{\dst^-}{\dst^+}\Dep(\pt_*\!)
\]
is indeed semi-singular. This vector is in fact a linear combination of
$w_{(2)}|_{q=1}$ and $w_{(8)}|_{n=0}$, and it is the vector presented in
Theorem~\ref{T:ss-q}(iii).\\

{\bf Case}  (3):
In order to evaluate $e_0^+w_{(3)}$
let us calculate $e_0^+ D_1D_2\,\dst^-_1\!\Dep\!(\pt_*\!)\pt_3$ first.
According to Proposition~\ref{P:eD} we get the following four terms:
\begin{eqnarray*}
e_0^+ D_1D_2\,\dst^-_1\!\Dep\!(\pt_*\!)\pt_3 &=&
D_1\{2\}D_2\{2\}\,e_0^+\dst^-_1\!\Dep\!(\pt_*\!)\pt_3
-
D_2\{2\}\dst^+_{3}B\dst^-_1\!\Dep\!(\pt_*\!)\pt_3\\
& &-
D_1\{2\}\dst^+_{3}f_2\dst^-_1\!\Dep\!(\pt_*\!)\pt_3+
\dst^+_{3}K\dst^-_1\!\Dep\!(\pt_*\!)\pt_3 \,.
\end{eqnarray*}
Now it is practical to compute these terms $\bmod \sdeg (<2)$.
\begin{eqnarray*}
D_1\{2\}D_2\{2\}\,e_0^+\dst^-_1\!\Dep\!(\pt_*\!)\pt_3
&\equiv&\,\,3\,
\dhind 2(\dhind 1 \Dep(\pt_*\!)\pt_2-\dhind 2 \Dep(\pt_*\!)\pt_1)\,,\\
D_2\{2\}\dst^+_{3}B\dst^-_1\!\Dep\!(\pt_*\!)\pt_3
&\equiv&\,\,\,\,
\dhind 2(\dhind 1 \Dep(\pt_*\!)\pt_2-\dhind 2 \Dep(\pt_*\!)\pt_1)\,,\\
D_1\{2\}\dst^+_{3}f_2\dst^-_1\!\Dep\!(\pt_*\!)\pt_3
&\equiv&-3\,
\dhind 2(\dhind 1 \Dep(\pt_*\!)\pt_2-\dhind 2 \Dep(\pt_*\!)\pt_1)\,,\\
\dst^+_{3}K\dst^-_1\!\Dep\!(\pt_*\!)\pt_3 \,
&\equiv& \,\,2\,
\dhind 2(\dhind 1 \Dep(\pt_*\!)\pt_2-\dhind 2 \Dep(\pt_*\!)\pt_1)\,.
\end{eqnarray*}
Looking at the whole expression for $w_{(3)}$ we immediately conclude
that
\begin{equation*}
 e_0^+w_{(3)} \equiv e_0^+ D_1D_2\,\dst^-_1\!\Dep\!(\pt_*\!)\pt_3
\equiv 3\,\dhind 2(\dhind 1 \Dep(\pt_*\!)\pt_2-\dhind 2
\Dep(\pt_*\!)\pt_1)
\quad \bmod \sdeg (<2)\,.
\end{equation*}
Thus $e_0^+w_{(3)}\neq 0 $, and
 we get no semi-singular vector in this case.\\

{\bf Cases}  (4)-(6): Each case here gives a  semi-singular vector.
They coincide with those of Theorem~\ref{T:ss-q}(ii), (i) and (iv).\\

{\bf Cases}  (7) and (10): It is straightforward to check that
\begin{equation*}
e_0^+w_{(7)}= - \dhind 2\dst^+_{3}\Dep(\pt_*\!) - \dst^-_{3}\dst^+_{312}\pt_2\,.
\end{equation*}
At the same time for $ w'_{(10)}=w_{(10)}|_{n=1}$ we get
\begin{equation*}
e_0^+w'_{(10)}= 3\, \dhind 2\dst^+_{3}\Dep(\pt_*\!) + 3\, \dst^-_{3}\dst^+_{312}\pt_2\,.
\end{equation*}
Clearly a linear combination $w=\tfrac{1}{3}w_{(7)}+ w'_{(10)}$ gives the
semi-singular vector. Up to notations this is the vector from
Theorem~\ref{T:ss-q}(v). And $ w=w_{(10)}|_{n=2}$ is a semi-singular
vector given in Theorem~\ref{T:ss-q}(vi).
\\

{\bf Cases}  (8) and (9): Here
\begin{eqnarray*}
e_0^+w_{(8)}&=& D_3^n e_0^+ \cycl{\dst^-}{\dst^-}{\dst^+}\Dep(\pt_*)
              - n \,D_3^{n-1}
              \dst^+_{3}\cycl{\dst^-}{\dst^-}{\dst^+}\Dep(\pt_*\!)
    - n \,D_3^{n-1}e_0^+ \dst^-_{123}\dst^+_{123}\pt_3 \\
   & & \qquad \qquad\qquad
    +\,\, n(n-1) D_3^{n-2}\dst^+_{3}\dst^-_{123}\dst^+_{123}\pt_3 \\
   &\equiv& n \dhind 3^{n-1}\left(
\dhind 1 (\dst^-_\circ \dst^+_\circ )_{13}+
\dhind 2 (\dst^-_\circ \dst^+_\circ)_{23}+ \dhind 3\dst^-_{3}\dst^+_{3}
\right)
\Dep(\pt_*\!) \s \bmod \,\sdeg(< n)\,.
\end{eqnarray*}
This gives no semi-simgular vector. And for $w_{(9)}$ we get
\begin{eqnarray*}
e_0^+w_{(9)}&=& D_1\{2\}\,e_0^+\dst^-_{12}\dst^+_{1}\Dep(\pt_*\!)
               - \dst^+_{3} B \dst^-_{12}\dst^+_{1}\Dep(\pt_*\!)
               -2\,e_0^+\dst^-_{312}\dst^+_{312}\pt_3 \\
       \equiv &-&\!\!\!\!\!\left(
\dhind 1(\dst^-_{1}\dst^+_{3} + 2\dst^-_{3}\dst^+_{1})+
\dhind 2(\dst^-_{2}\dst^+_{3} + 2\dst^-_{3}\dst^+_{2})+
4\,\dhind 3\,\dst^-_{3}\dst^+_{3}  \right)
\Dep(\pt_*\!) \, \bmod \sdeg(< 1).
\end{eqnarray*}
We see that no linear combination of $w_{(9)}$ and $w_{(8)}|_{n=1}$
could provide a semi-simgular vector.
{\flushright \epf \\
}

\section{From semi-singular to singular vectors.}
\label{sec:8}

The results obtained in Sections~\ref{sec:3}-\ref{sec:7} and
Remark~\ref{R:ass} permit us to write explicitely not only admissible,
but all semi-singular vectors, as it is done in the propositions
below. Then we begin to work on determining
the singular vectors following the directions
outlined in Section~\ref{sec:2}.
\medskip

Let us have in mind that the
grading of $L_-$ extends to the grading on $U(L_-)\otimes_\CC {V}$,
where we suppose grading on $V$ being trivial. Hence for an element
$w$ in $U(L_-)\otimes_\CC {V}$ it is natural to define a
 degree $\udeg w$ to be equal to the grading {\it with the opposite sign.}

\begin{Proposition}
\label{P:ss-p}
Let $V$ be an irreducible $s\ell(3)$-submodule in $\CC[x_1,x_2,x_3 ]$
generated by $x_1^p$. Then semi-singular
$s\ell(2)$-weight vectors $w$ in $U(L_-)\otimes_\CC {V}$
are the following (up to linear combinations).
\begin{align*}
\s\udeg w=0: \s\,&    x_1^p\,,                  \\
\s\udeg w=1: \s\,&   \dst^+_1x_1^p\,, & &
                              \dst^-_1x_1^p\,,\\
\s\udeg w=2: \s\,&    \dst^-_1\dst^+_1x_1^p  \,,\\
\s\udeg w=3: \s\,&   a=\dst^+_{123}\,, & & b=f_3a=
\cycl{\dst^-}{\dst^+}{\dst^+}-\hat{\De}(\dst^+_*)\,,\s & \\
 \,& && c=\tfrac{1}{2}f_3b=\cycl{\dst^-}{\dst^-}{\dst^+}-\hat{\De}(\dst^-_*)\,,
   \s \s  d=\dst^-_{123}\,,\\
\s\udeg w=4: \s\,& \dst^-_1a\,,&&
       \dst^-_1b\,, \s\s \dst^-_1c \,,\\
\s\udeg w=6: \s\,&
\hat{\De}(\dst^-_*)\,a\,, && \hat{\De}(\dst^-_*)\,b
\s\s \hat{\De}(\dst^-_*)\,c\,,\\
 &\s\s\text{and} &&
d\,a - a\,d= 2\,d\,a + \hat{\De}(\dst^-_*)\,b\,,\\
\s\udeg w=7: \s\,&
\dst^-_1\hat{\De}(\dst^-_*)a\,, && \dst^-_1\hat{\De}(\dst^-_*)b.
\end{align*}
\end{Proposition}

Let us fix and keep for the following the above the notations $a,\,b,\,c,\,d$
for the elements of $U(L_-)$ (these notations were also used in [4]).

\begin{Remark}\label{R:abcd}
{\sl
One should have in mind that $a,\,b,\,c,\,d$ make a standard weight
vector basic for a four-dimensional $s\ell (2)$-representation with
the relations
\begin{eqnarray}
  \label{eq:8.1}
f_3\,a=b\,,\s  f_3\,b=2c\,,\s f_3\,c=3d\, \\
e_3\,b=3a\,,\s e_3\,c=2b\,, \s e_3\,d=c\,, \notag
\end{eqnarray}
and that for any $i=1,2,3$
\begin{eqnarray}
  \label{eq:8.2}
\,\,\dst^+_i a=0\,,\,
\dst^+_i b=-\dst^-_i a\,,\,
\dst^+_i c=-\dst^-_i b\,,\,
\dst^+_i d=-\dst^-_i c\,,\,
           0=\dst^-_i d\,.
\end{eqnarray}
Clearly the same rules are true for the multiplication on the other
side. Also
\begin{eqnarray}
  \label{eq:8.3}
 &\, & a\,\Dep(\pt_*)=0\,,\,   \, \notag \\
& a\,\Dem(\pt_*)=-b\,\Dep(\pt_*)\,,
& b\,\Dem(\pt_*)=-c\,\Dep(\pt_*)\,,
\s c\,\Dem(\pt_*)=-d\,\Dep(\pt_*)\,,\\
&\, & d\,\Dem(\pt_*)=0\,, \,\notag
\end{eqnarray}
and the same is true for $\hat{\De}(\dst^-_*)$, $\hat{\De}(\dst^+_*)$,
and for the multiplication from the other side as well.
}
\end{Remark}

\begin{proof} We are to combine the vectors of Theorem~\ref{T:ss-p} with
those from Corollary~\ref{C:qs-deg0-p} that are semi-singular,
apply $\ph$ according to Remark~\ref{R:ass}. In fact the notations
of the proposition are written in a way
is more convenient for the future calculations, namely the
vectors are such
that they are weight vectors and make the bases for
irreducible $s\ell(2)$-representations.
In particular from the relations
\begin{align}\label{eq:a-d}
a\,b &= 0, & b\,a &= 0, & a\,c &= c\,a, & b\,c - c\,b &= 3(a\,d - d\,a),\\
b\,d &= 0, & d\,b &= 0, & b\,d &= d\,b, &
a\,d + d\,a &= - \hat{\De}(\dst^-_*)\,b, \notag
\end{align}
it follows that $f_3(a\,d - d\,a)=e_3(a\,d - d\,a)=0$ and thus
the vector
$d\,a - a\,d = 2\,d\,a + \hat{\De}(\dst^-_*)\,b$
makes an one-dimensional $s\ell(2)$-representation.
We leave the rest for the reader.
\end{proof}

Similarly Theorem~\ref{T:ss-q} and Corollary~\ref{C:ss-deg0-q}
lead us to the following proposition.
\begin{Proposition}
\label{P:ss-q}
Let $V$ be an irreducible $s\ell(3)$-submodule in
$\CC[\pt_1,\pt_2,\pt_3 ]$
generated by $\pt_3^q$. The semi-singular
$s\ell(2)$-weight vectors $w$ in $U(L_-)\otimes_\CC {V}$,
up to linear combinations, are:
\begin{align*}
\s\udeg w=0: \s\,&    \pt_3^q\,,  & &                \\
\s\udeg w=1: \s\,&  \Dep(\pt_*)\pt_3^{q-1}\,, & & \Dem(\pt_*)\pt_3^{q-1}
                                              \,,\\
\s\udeg w=2: \s\,&  \dst^+_1\Dep(\pt_*)\,,
                && \dst^-_1\Dep(\pt_*)\,, \s
                   \dst^-_1\Dem(\pt_*)\,,  \\
 &  && \hspace{-15ex}\dhind 2\pt_2-\dhind 3\pt_3-
\Dem(\dst^+_1\pt_*)= \dst^+_1\Dem(\pt_*), \s\s\text{ and }\s
\mbox{$
\Dem(\Dep(\pt_*)\,\pt_*)$}\,, & &\, \\
\s\udeg w=3: \s\,&   \dst^-_1  \dst^+_1\Dep(\pt_*)\,,
  && \\
\s\udeg w=4: \s\,& b\, \Dep(\pt_*) \,,
  && c\,\Dep(\pt_*)\,, \s\s  d\,\Dep(\pt_*)\,, \s\text{ and } \\
 & && \hspace{-25ex}
\mbox{$
\dst^-_1\dst^+_1\Dem(\Dep(\pt_*)\pt_*)=
\dhind 2\dst^-_1\Dep(\pt_*)\pt_2-\dhind 3\dst^-_1\Dep(\pt_*)\pt_3-
\dst^-_1\Dem(\dst^+_1\Dep(\pt_*)\pt_*)
$}\,, & &\, \\
\s\udeg w=5: \s\,&
\dst^-_1\,a\,\Dem(\pt_*)\,, &&
\dst^-_1\,b\,\Dem(\pt_*)\,,
\\
\s\udeg w=7: \s\,& \hat{\De}(\dst^-_*)\,b\,\Dep(\pt_*)\,, &&
\hat{\De}(\dst^-_*)\,c\,\Dep(\pt_*)\,.
\end{align*}
\end{Proposition}
\begin{proof}
Let us notice that the vectors with $\udeg w\geq 4$ in the proposition
are  not all written in the $\La^-\La^+$ order.  For the
proof it is better to rewrite them in this order, namely
\begin{eqnarray}
 b\, \Dep(\pt_*) \label{eq:8.4}
&=&(\cycl{\dst^-}{\dst^+}{\dst^+}-\hat{\De}(\dst^+_*))\, \Dep(\pt_*) \\
&=&\Dem(\dst^+_{123}\pt_*\!)-\hat{\De}(\dst^+_*)\, \Dep(\pt_*)\notag\\
&=&
-(\dhind 1 \dst^+_1 +\dhind 2 \dst^+_2 +
\dhind 3 \dst^+_3)\Dep(\pt_*)+\Dem(\dst^+_{123}\pt_*\!)\,, \notag\\
 c\, \Dep(\pt_*) \label{eq:8.5}
&=&(\cycl{\dst^-}{\dst^-}{\dst^+}-\hat{\De}(\dst^-_*))\, \Dep(\pt_*)\,, \\
\dst^-_1\,a\,\Dem(\pt_*)&=&-\dst^-_1\,b\,\Dep(\pt_*)\label{eq:8.6}\\
&=&\notag
\dst^-_1\,(\hat{\De}(\dst^+_*)-\cycl{\dst^-}{\dst^+}{\dst^+})\, \Dep(\pt_*)\,,
\\
  \label{eq:8.7}
\hat{\De}(\dst^-_*)\,b\,\Dep(\pt_*)&=&
\hat{\De}(\dst^-_*)\,(\cycl{\dst^-}{\dst^+}{\dst^+}-\hat{\De}(\dst^+_*)
)\,\Dep(\pt_*)\\
&=&
-\tfrac{1}{12}\left(
D_1^2\,\dst^-_1\dst^+_1\Dep(\pt_*)
-6\,D_1\,\dst^-_1\Dem(\dst^+_{123}\pt_*)\right)\,. \notag
\end{eqnarray}
We see that the proposition follows  pretty straightforwardly
from Theorem~\ref{T:ss-q}, Corollary~\ref{C:ss-deg0-q}
and Remark~\ref{R:ass}.
\end{proof}

\medskip
Let us recall that we study the highest singular vectors $\mx{\bf v}$ in the
generalized Verma modules $M(\mx{\bf V})$,
$\mx{\bf  V}=V\otimes T$ where
$V$ is an $s\ell(3)\oplus g\ell(1)$-module and $T$ is an
$s\ell(2)$-module.
Practically it is more convenient to consider both $V$ and
$T$ being $\fg_0$-modules with the actions of $s\ell(2)$ on $V$
and $s\ell(3)$ on $T$ being trivial and some action of $g\ell(1)$
on both.
The tensor products in
\eqref{eq:VTiso} are then the tensor products of $\fg_0$-modules.

As it was said in
Remark~\ref{R:p-or-q}
we suppose that
$V$ is an irreducible $\fg_0$-submodule  of either
$\CC[x_1,x_2,x_3 ]$ or $\CC[\,\dpind1, \dpind2, \dpind3]$
with the action of $g\ell(1)$ naturally defined by Remark~\ref{R:gweit}.
For the module $T$ we take an $(r+1)$-dimensional module with
the basic $\{ z_+^{r-j}z_-^{j}[\tht] \,|\, j=0,1,\ldots,r\,\}$ and the
$\fg_0$-module structure given by the rules
\begin{eqnarray}
  \label{eq:8.8}
 e_3\, z_+^{r-j}z_-^{j}[\tht] &=&j\,z_+^{r-j+1}z_-^{j-1}[\tht]\,,\\\notag
 f_3\, z_+^{r-j}z_-^{j}[\tht] &=&(r-j)z_+^{r-j-1}z_-^{j+1}[\tht]\,,\\
   Y\, z_+^{r-j}z_-^{j}[\tht] &=& (\tht-r)\,z_+^{r-j}z_-^{j}[\tht]      \label{eq:Y-T}
\end{eqnarray}
where $\tht \in \CC$ could be arbitrary,
and as we said
$s\ell(3)$ acts trivially. \\
{ To remind} the parameters of $T$ {\it we write
$T=T(r,\tht)$} when it seems needed.
\medskip

We shall
use for $\mx{\bf v}\in M(\mx{\bf V})=M(V)\otimes T$ the decomposition
\begin{align}
  \label{eq:decomp}
\mx{\bf v}&=\sum_{j\leq j_0} u_j \cdot t_j ,\,
\end{align}
$\text{where }\, u_j\in M(V),
\,\,\,t_j = c_j\, z_+^{r-j}z_-^{j}[\tht]\in T\,$, and
$c_j\in \CC $.
Let $ \mx{\bf v}$ be the $\fg_0$-highest weight vector, then
it is necessary that $c_{0}\neq 0$, and $u_{0}$ is the $s\ell(2)$-highest
vector of a finite-dimensional $s\ell(2)$-submodule $S$ in $M(V)$.

Suppose $\mx{\bf v} $ is also singular. Then
from Proposition~\ref{P:2.3} it follows that
$u_j$ are the semi-singular weight vectors.
Therefore only vectors listed in
Propositions~\ref{P:ss-p},~\ref{P:ss-q}
are to be considered for $u_j$.
This is the key idea leading to the classification of singular vectors.
The results are summed up as the following main theorem.
\bTh \label{T:sing-v} The non-trivial $\fg_0$-highest singular vectors
$\mx{\bf v}$ are the following ones (the value of $\tht$ is written in the
square brackets):
\begin{tabbing}
\hspace*{1ex}\=\hspace{6ex}\=\hspace{10ex}\=\hspace{10ex}\=\hspace{10ex} \kill
\\
\>1a. \>$\mx{\bf v}=\dst^+_1x_1^p\cdot z_+^{r}[0]$, $\s(\,p,r\geq 0,\, q=0\,)$,\\
\\
\>1b. \>$\mx{\bf v}=
         \dst^+_1x_1^p\cdot z_+^{r-1}z_-[2r+2] -\dst^-_1x_1^p\cdot
         z_+^{r}[2r+2]$, $\s(\,p\geq 0,\,r\geq 1,\, q=0\,)$, \\
\\
\>2. \>$\mx{\bf v}=
         \dst^+_1\dst^-_1x_1^p\cdot 1[2]$, $\s(\,p\geq 0,\, q=r=0\,)$,\\
\\
\>3. \>$\mx{\bf v}=
          \dst^+_{123}\cdot z_+^{r}[-2]$, $\s(\,p=q=0,\,\,r\geq 0)$,\\
\\
\>4. \>$\mx{\bf v}=
                a\cdot z_+^{r-3}z_-^3[2r]
                  - b\cdot z_+^{r-2}z_-^2[2r]
                      + c\cdot z_+^{r-1} z_-[2r]
                          - d\cdot z_+^{r}[2r]$, \\
\> \> \hspace*{19em}         $\s(\,p=q=0,\, r\geq 3\,)$,\\
\\
\>5. \>$\mx{\bf v}=
            \dst^-_1a\cdot z_-^2[4] -  \dst^-_1b\cdot z_+z_-[4]
                 +  \dst^-_1c\cdot z_+^2[4]$,  $\s(\,p=q=0, \,r=2\,)$,\\
\\
\>6. \>$\mx{\bf v}=
           \hat{\De}(\dst^-_*)\,a\cdot z_-[0] -
                   (d\,a + \hat{\De}(\dst^-_*)\,b)\cdot z_+[0]$,
                                            $\s(\,p=q=0, \,r=1\,)$,\\
\\
\>7. \>$\mx{\bf v}=
         \Dep(\pt_*)\pt_3^{q-1}\cdot z_+^{r}[-2]$,
                         $\s(\,p=0,\, q\geq 1,\, r\geq 0\,)$,\\
\\
\>8. \>$\mx{\bf v}=
         \Dep(\pt_*)\pt_3^{q-1}\cdot z_+^{r-1}z_-[2r] -
            \Dem(\pt_*)\pt_3^{q-1}\cdot z_+^{r}[2r]$,
                                          $\s(\,p=0,\, q\geq 1,\,r\geq 1\,)$,\\
\\
\>9. \>$\mx{\bf v}=
        \dst^+_1\Dep(\pt_*)\cdot z_-[2]
         -  \dst^+_1\Dem(\pt_*)  \cdot z_+[2]$, $\s(\,p=0,\,q=1,\,r=1\,)$, \\

\\
\>10. \>$\mx{\bf v}=
                  \Dem(\Dep(\pt_*)\pt_*)\pt_3^{q-2}\cdot 1[0]$,
                                           $\s(\,p=0,\,q\geq 2,\,r=0\,)$, \\

\\
\>11. \>$\mx{\bf v}=
                  b\Dep(\pt_*)\cdot 1[0]$, $\s(\,p=0,\,q=1, \,r=0\,)$.
\end{tabbing}
\eTh
\begin{Remark}\label{R:8.5}
{\sl
Let us compare the list above with the results presented
in [4]. In the notations of [4], Theorem~1.2, vectors given in (1a)
are of type $A$, i.e. belong to $M(p,0;r;y_A)$, where
$y_A=\tfrac{2}{3}p-r$.
We only need to notice that
according to \eqref{eq:Y-T} and Remark~\ref{R:gweit}
\begin{equation*}
Y\,x_1^p\cdot z_+^{r}[\tht] = \tfrac{2}{3}p-r+\tht.
\end{equation*}
Similarly vectors from (1b) and (2) are of type $B$, i.e.
belong to $M(p,0;r;y_B)$, where $y_B=\tfrac{2}{3}p+r+2$.
But from
\begin{equation*}
Y\,\pt_3^q\cdot z_+^{r}[\tht] =-\tfrac{2}{3}q-r+\tht
\end{equation*}
it follows that
vectors (7) and (3) are of type $C$, i.e. are in
$M(0,q;r;y_C)$, where $y_C=-\tfrac{2}{3}q-r-2$, and
the rest, namely (8), (10), (4), (5), (6) and (9) are of type $D$, i.e.
belong to various modules $M(0,q;r;y_D)$, where $y_D=-\tfrac{2}{3}q+r$.

It is less straightforward to establish the correspondence between
explicit expressions for the singular vectors given in [4]
and the ones given above. First of all we notice
that in [4] we worked with the two
versions of a 2-dimensional irreducible representation of
$s\ell(2)$, one having the basis $\{ z_+,\,  z_-\}$, the other
with the basis $\{-\pt_-,\,\pt_+\}$. Now we need the same spaces, but
with the action extended
to $s\ell(2)\oplus Y$, therefore the bases are to be written as
$\{ z_+[\tht],\,  z_+[\tht]\}$,
$\{-\pt_-[\tht'],\,\pt_+[\tht']\}$, and the representations are isomorphic
if and only if $\tht'=\tht+2$. With this in mind
%
we have no problem indentifying the ``general'' $A,B,C,D$ cases
(1a), (1b), (7), (8) with those of Corollary~2.5 in [4].

Further, our vectors in (2) correspond to those of the case (a) of
Corollary~2.8 in [4], and our vectors in (3) correspond to the case (c),
in (4) to the case (d) and in (10) to the case (b).

The vector $w_1$ (given by (2.12) of [4]) is nothing but
the vector from (11), up to a sign. The vector $w_2$, (defined by
(2.16) in [4]) is the vector from (5).
The vector $w_3$ (see (2.18) and Lemma~5.26  of [4])
is the vector from (6),
and  the vector $w_4$ (given by (2.20) of [4]) is from (9).
Thus we have a one-one correspondence between vectors
listed in Theorem~\ref{T:sing-v} above and in Theorem~2.10 of [4].
}
\end{Remark}
\begin{proof}
It is clear that Propositions~\ref{P:ss-p},~\ref{P:ss-q}
provide us with few choices for the $s\ell(2)$-module $S$
containing the coefficients $u_{j}$  of \eqref{eq:decomp}.
Clearly we can suppose that $S$ is irreducible. Because
there are only modules $S$ with $\dim S \leq 4$ it is
easy to write explicitely possible
$s\ell(2)$-highest candidates $\mx{\bf v}$ for the
singular vectors. It follows from the propositions that
the vectors will be semi-singular. Such a vector
will be singular if and only if it will satisfy
the condition $e_0\,\mx{\bf v}=0$.

We shall
study the possible choices and check when the condition
is satisfied. Let us remind that
\begin{align}\label{eq:e0-rel}
{}[e_0,\,\dst^-_1]&= -2f_3   , &{}[e_0,\,\dst^+_1]&= \,h_0    , \\\notag
{}[e_0,\,\dst^-_2]&= \,0   , &{}[e_0,\,\dst^+_2]&= \,f_1      , \\
{}[e_0,\,\dst^-_3]&= \,0   , &{}[e_0,\,\dst^+_3]&= -f_{12}   , \notag
\end{align}
and $[e_0,\,\hpt_1]=0$,  $[e_0,\,\hpt_2]=\dst^-_3 $, $[e_0,\,\hpt_3]= -\dst^-_2$. \\

1) $\,S=\lsp{\,\dst^+_1x_1^p\,,  \dst^-_1x_1^p\,}$. \\
Here either $\mx{\bf v}=\dst^+_1x_1^p\cdot z_+^{r}[\tht]$,
or
$
\mx{\bf v}= \dst^+_1x_1^p\cdot z_+^{r-1}z_-[\tht] -\dst^-_1x_1^p\cdot
         z_+^{r}[\tht].
$
\hspace{0em}From the condition $e_0\,\mx{\bf v}=0$ it follows
$\tht=0$ for the former which gives us vectors from ({\it 1a}) .
For the latter, we compute
\begin{eqnarray*}
e_0\mx{\bf v}
&=&e_0(\dst^+_1x_1^p\cdot z_+^{r-1}z_-[\tht] -\dst^-_1x_1^p\cdot z_+^{r}[\tht]\,)\\
   &=&h_0\,x_1^p\cdot z_+^{r-1}z_-[\tht]+2f_3\,x_1^p\cdot z_+^{r}[\tht]\\
   &=&(2-\tht)x_1^p\cdot z_+^{r-1}z_-[\tht]+2r\,x_1^p\cdot z_+^{r-1}z_-[\tht]\\
   &=&(2-\tht+2r)x_1^p\cdot z_+^{r-1}z_-[\tht],
\end{eqnarray*}
hence $\tht=2+2r$
which gives us vectors from
({\it 1b}) of the theorem. \\

2) $\,S=\lsp{\,\dst^-_1\dst^+_1x_1^p\,}$. \\
We come to $\mx{\bf v}=\dst^-_1\dst^+_1x_1^p\cdot z_+^{r}[\tht]$.
Now
\begin{eqnarray*}
e_0\,\dst^-_1\dst^+_1x_1^p\cdot z_+^{r}[\tht]&=&
(-2f_3)\dst^+_1x_1^p\cdot z_+^{r}[\tht]
-\dst^-_1(h_0\,x_1^p\cdot z_+^{r}[\tht])\\
& =&
-2\dst^-_1x_1^p\cdot z_+^{r}[\tht]-2r\dst^+_1x_1^p\cdot z_+^{r-1}z_-[\tht]
+\tht\,\dst^-_1x_1^p\cdot z_+^{r}[\tht],
\end{eqnarray*}
hence $\tht=0$, $r=0$. We come to the vector from ({\it 2}).\\

3) $\,S=\lsp{\,a,\,b,\,c,\,d\, }$ (in the notations of Proposition~\ref{P:ss-p}). \\
We have four choices for $\mx{\bf v}$.
\begin{eqnarray*}
\mx{\bf v}_1&=&a\cdot z_+^{r}[\tht],\\
\mx{\bf v}_2&=&b\cdot z_+^{r}[\tht]
               -3a \cdot z_+^{r-1}z_-[\tht],\\
\mx{\bf v}_3&=&c\cdot z_+^{r}[\tht]
               -2b \cdot z_+^{r-1}z_-[\tht]
                +3a \cdot z_+^{r-2}z_-^{2}[\tht],\\
\mx{\bf v}_4&=&d\cdot z_+^{r}[\tht]
                -c\cdot z_+^{r-1}z_-[\tht]
                  +b\cdot z_+^{r-2}z_-^{2}[\tht]
                    -a\cdot z_+^{r-3}z_-^{3}[\tht].
\end{eqnarray*}
\hspace{0em}From \eqref{eq:e0-rel} it follows
\begin{align}
\label{eq:e0-a}
[e_0,\,a] \,=\, e_0\,\dst^+_1\dst^+_2\dst^+_3 &= h_0 \dst^+_2\dst^+_3 -
               \dst^+_1 f_1 \dst^+_3 +  \dst^+_1\dst^+_2 f_{12} \\
       &= \dst^+_2\dst^+_3 (h_0 - 2)- \dst^+_1\dst^+_3f_1 +
               \dst^+_1\dst^+_2 f_{12}. \notag
\end{align}
We get immediately that $e_0\,\mx{\bf v}_1=(-\tht-2)
\dst^+_2\dst^+_3\cdot z_+^{r}[\tht]$,
hence $\tht=-2$ and we come to the vector from ({\it 3}).

Notice that
${}[e_0,\,f_3]= 0$. 
This makes it easy to calculate from \eqref{eq:e0-a} that
\begin{align}
\label{eq:e0-b}
[e_0,\,b] &= (\dst^-_2\dst^+_3 + \dst^+_2\dst^-_3)(h_0-2) +
                  \dst^+_2\dst^+_3 (-2f_3)
    -  (\dst^-_1\dst^+_3 + \dst^+_1\dst^+_3)f_1 + \\
    & \qquad\qquad\qquad\qquad
        + (\dst^-_1\dst^+_2 + \dst^+_1\dst^-_2)f_{12}. \notag
\end{align}
Therefore
\begin{align*}
e_0\,\mx{\bf v}_2=&\,\,
(-\tht-2)(\dst^-_2\dst^+_3 + \dst^+_2\dst^-_3)\cdot z_+^{r}[\tht]+
(-2r + 3\tht)\dst^+_2\dst^+_3 \cdot z_+^{r-1}z_-[\tht].
\end{align*}
We come to equations $\tht=-2$ and $2r=3\tht$ that gives $r=-3$ which
is impossible.

Similarly
\begin{align}
\label{eq:e0-c}
[e_0,\,c] &= \dst^-_2\dst^-_3(h_0 -2)+
                      (\dst^-_2\dst^+_3 +\dst^+_2\dst^-_3)(-2f_3)
 - \dst^-_1\dst^-_3f_1 + \dst^-_1\dst^-_2f_{12}.
\end{align}
Formulae  \eqref{eq:e0-a}, \eqref{eq:e0-b} and \eqref{eq:e0-c}
imply that
\begin{align*}
e_0\,\mx{\bf v}_3\, =&\,\,(-\tht-2)\dst^-_2\dst^-_3\cdot z_+^{r}[\tht]
\qquad\qquad\qquad \qquad\qquad \qquad \\
&+(-2r + 3\tht)(\dst^-_2\dst^+_3 + \dst^+_2\dst^-_3)\cdot z_+^{r-1}z_-[\tht]\\
&+(4(r-1)r + 3(2-\tht))\dst^+_2\dst^+_3 \cdot z_+^{r-2}z_-^2[\tht].
\end{align*}
We see that it should be $\tht=2$ and $r=\tht$ which give the
impossible value $r=-2$.

For $d$ we immediately get from \eqref{eq:e0-rel}
\begin{align}
\label{eq:e0-d}
[e_0,\,d] &= \dst^-_2\dst^-_3(-2f_3).\qquad\qquad \qquad\qquad \qquad\qquad\qquad \qquad
\end{align}
Therefore
\begin{align*}
e_0\,\mx{\bf v}_4\, =
&\,\,(-2r+\tht)\dst^-_2\dst^-_3\cdot z_+^{r-1}z_-[\tht]
+(2(r-1)+(2-\tht))(\dst^-_2\dst^+_3 +\dst^+_2\dst^-_3)\cdot z_+^{r-2}z_-^{2}[\tht]\\
&+(-2(r-2)-(4-\tht))\dst^+_2\dst^+_3\cdot z_+^{r-3}z_-^{3}[\tht].
\end{align*}
We conclude that $ \tht= 2r$, $r\geq 3$ and this is what stated in
({\it 4}).\\

4) $\,S=\lsp{\,\dst^-_1a,\,\dst^-_1b,\,\dst^-_1c\, }$. \\
\hspace{0em}From \eqref{eq:8.1},  \eqref{eq:8.2} we conclude that there are three
choices for $\mx{\bf v}$.
\begin{eqnarray*}
\mx{\bf v}_1&=& \dst^-_1a\cdot z_+^{r}[\tht],\\
\mx{\bf v}_2&=&\dst^-_1b\cdot z_+^{r}[\tht]
               -2\dst^-_1a \cdot z_+^{r-1}z_-[\tht],\\
\mx{\bf v}_3&=&\dst^-_1c\cdot z_+^{r}[\tht]
               -\dst^-_1b \cdot z_+^{r-1}z_-[\tht]
                +\dst^-_1a \cdot z_+^{r-2}z_-^{2}[\tht].
\end{eqnarray*}
On the other hand $[e_0,\,\dst^-_1x]= (-2f_3)x - \dst^-_1[e_0,\,x]$.
Then using \eqref{eq:e0-a} we easily calculate that
\begin{align*}
e_0\,\mx{\bf v}_1&=\,(-2f_3)a\cdot z_+^{r}[\tht]
                       - \dst^-_1(e_0\,a\cdot z_+^{r}[\tht])\\
&=-2b\cdot z_+^{r}[\tht]-2r\,a\cdot z_+^{r-1}z_-[\tht]-
(-\tht-2)\dst^-_1\dst^+_2\dst^+_3 \cdot z_+^{r}[\tht] \neq 0 .
\end{align*}
Thus $\mx{\bf v}_1 $ is never singular.

Similarly for $\mx{\bf v}_2$ we get
\begin{align*}
e_0\,\mx{\bf v}_2 =&\,
(-4)c\cdot z_+^{r}[\tht]
+(-2r+4)b\cdot z_+^{r-1}z_-[\tht]
+4(r-1)a\cdot z_+^{r-2}z_-^{2}[\tht]\\
&+(\tht+2)
(\dst^-_1\dst^-_2\dst^+_3 + \dst^-_1\dst^+_2\dst^-_3)\cdot z_+^{r}[\tht]+
(2r - 2\tht)\dst^-_1\dst^+_2\dst^+_3 \cdot z_+^{r-1}z_-[\tht].
\end{align*}
Also it is never zero, we get no singular vector.

For the last one
\begin{align*}
e_0\,\mx{\bf v}_3\, =&\,\,
(\tht-4)d\cdot z_+^{r}[\tht]
-2(r-2)(c\cdot z_+^{r-1}z_-[\tht]-b\cdot z_+^{r-2}z_-^{2}[\tht]
         +a\cdot z_+^{r-3}z_-^{3}[\tht])
\qquad\qquad\qquad  \\
&+(2r -\tht)(\dst^-_1\dst^-_2\dst^+_3
+\dst^-_1 \dst^+_2\dst^-_3)\cdot z_+^{r-1}z_-[\tht]
+(-2r + \tht)\dst^-_1\dst^+_2\dst^+_3 \cdot z_+^{r-2}z_-^2[\tht].
\end{align*}
We get zero only when $r=2$, $\tht=4$ which gives us the singular vector
from ({\it 5}).\\

5) $\,S=\lsp{\hat{\De}(\dst^-_*)a,
           \,\hat{\De}(\dst^-_*)b,
           \,\hat{\De}(\dst^-_*)c}\oplus
        \lsp{d\,a - a\,d}$. \\
We have here the following choices for $\mx{\bf v}$.
\begin{eqnarray*}
\mx{\bf v}_1&=& \hat{\De}(\dst^-_*)a\cdot z_+^{r}[\tht],\\
\mx{\bf v}_2&=&\hat{\De}(\dst^-_*)(b\cdot z_+^{r}[\tht]
               -2a\cdot z_+^{r-1}z_-[\tht])\,,\\
\mx{\bf v}_3&=&(d\,a - a\,d)\cdot z_+^{r}[\tht]
=(2\,d\,a + \hat{\De}(\dst^-_*)\,b)\cdot z_+^{r}[\tht]\,,\\
\mx{\bf v}_4&=&\hat{\De}(\dst^-_*)(c\cdot z_+^{r}[\tht]
               - b\cdot z_+^{r-1}z_-[\tht]
                + a\cdot z_+^{r-2}z_-^{2}[\tht])\,,
\end{eqnarray*}
where $\mx{\bf v}_2$ and $\mx{\bf v}_3$ have the same weight, so we
have check their linear combinations as well.

First of all
\begin{eqnarray}
  \label{eq:e0-D_}
e_0\,\hat{\De}(\dst^-_*)&=& -2\,\dst^-_2\dst^-_3-2\,\dhind 1\,f_3
- \hat{\De}(\dst^-_*)\,e_0\,.
\end{eqnarray}
Now from \eqref{eq:e0-a} we get
\begin{align*}
e_0\,\mx{\bf v}_1 =&\,
-2\,\dst^-_2\dst^-_3\,a \cdot z_+^{r}[\tht]
-2\dhind 1( b\cdot z_+^{r}[\tht]
-r\,a \cdot z_+^{r-1}z_-[\tht])\\
&\qquad+(\tht+2)\hat{\De}(\dst^-_*)\dst^+_2\dst^+_3 \cdot z_+^{r}[\tht]\,,
\end{align*}
that is always non-zero.

Now we shall calculate the left hand side of $e_0\,\mx{\bf v}_2=0$.
Similarly
from \eqref{eq:e0-D_} we conclude that the only term
containing $z_+^{r-2}z_-^{2}[\tht]$
is $4(r-1)\dhind 1 a\cdot z_+^{r-2}z_-^{2}[\tht]$.
At the same time
\begin{eqnarray}
  \label{eq:e0-da}
e_0\,da \cdot z_+^{r}[\tht]  &=&
\dst^-_2\dst^-_3 \,(-2f_3)a \cdot z_+^{r}[\tht]
-d\, e_0 a\cdot z_+^{r}[\tht]
\notag \\
  &=&
-2 \dst^-_2\dst^-_3 b \cdot z_+^{r}[\tht]
  -2r\dst^-_2\dst^-_3a  \cdot z_+^{r-1}z_-[\tht]
+(\tht+2)\,d\, \dst^+_2\dst^+_3\cdot z_+^{r}[\tht]\,.\notag
\end{eqnarray}
This implies
that no term in $e_0\,\mx{\bf v}_3$
contains $z_+^{r-2}z_-^{2}[\tht]$.
Therefore unless $r=1$ we have got
$ e_0(\al\mx{\bf v}_2 + \be\mx{\bf v}_3) \neq 0$.

Now let $r=1$, then
\begin{align*}
e_0\,\hat{\De}(\dst^-_*)a\cdot z_-[\tht]=&
-2\dst^-_2\dst^-_3 \,a\cdot z_-[\tht]
-2\dhind 1 b\cdot z_-[\tht]\\
&\qquad+\tht\hat{\De}(\dst^-_*)\dst^+_2\dst^+_3\cdot z_-[\tht]\,,\\
e_0\,\hat{\De}(\dst^-_*)b\cdot z_+[\tht]=&
-2\dst^-_2\dst^-_3 \,b\cdot z_+[\tht]
-2\dhind 1 b\cdot z_-[\tht]
-4\dhind 1 c\cdot z_+[\tht]\\
&\qquad
+\tht\hat{\De}(\dst^-_*)(\dst^-_2\dst^+_3
+ \dst^+_2\dst^-_3)\cdot z_+[\tht]
+2\hat{\De}(\dst^-_*)\dst^+_2\dst^+_3\cdot z_-[\tht]\,.
\end{align*}
It is not difficult to conclude that only when $\tht=2$
the vector $\mx{\bf v}= \mx{\bf v}_2 + \mx{\bf v}_3$ could be
singular.
This vector is indeed singular
(equal to $w_3$ in the notations of [4]),
and this leads us to the case ({\it 6}) of the theorem.\\

6) $\,S=\lsp{\dst^-_1\hat{\De}(\dst^-_*)a,
           \,\dst^-_1\hat{\De}(\dst^-_*)b}$. \\
We have to work with vectors
\begin{eqnarray*}
\mx{\bf v}_1&=& \dst^-_1\hat{\De}(\dst^-_*)a\cdot z_+^{r}[\tht],\\
\mx{\bf v}_2&=&\dst^-_1\hat{\De}(\dst^-_*)(b\cdot z_+^{r}[\tht]
               -a\cdot z_+^{r-1}z_-[\tht])\,.
\end{eqnarray*}

For $\mx{\bf v}_1$ it is clear that
\begin{eqnarray*}
e_0\mx{\bf v}_1&=& (\ldots)\cdot z_+^{r}[\tht]
   -2r(\dst^-_2\dst^+_3 + \dst^+_2\dst^-_3)a  \cdot z_+^{r-1}z_-[\tht]\,,
\end{eqnarray*}
hence we are left with $r=0$. Then
\begin{eqnarray*}
e_0\mx{\bf v}_1&=&
(2da -2\hat{\De}(\dst^-_*)b+2\dst^-_1\dhind 1  b
-(\tht+2)\dst^-_1\hat{\De}(\dst^-_*)(\dst^-_2\dst^+_3 + \dst^+_2\dst^-_3)
)\cdot z_+[\tht]\,,
\end{eqnarray*}
that never gives zero ($da$ could not cancel with other terms).

For $\mx{\bf v}_2$ again we can first compute the coefficient
by $ z_+^{r-2}z_-^{2}[\tht]$. The computation gives
$-2(r-1)(\dst^-_2\dst^+_3 + \dst^+_2\dst^-_3)a$ so $r=1$.
Then we can calculate the the coefficient by $ z_+^{r-1}z_-[\tht]$.
Here we get
\[
-da+(2r+2-\tht)\dst^-_1\hat{\De}(\dst^-_*)\dst^+_2\dst^+_3 \neq 0\,,
\]
hence $\mx{\bf v}_2$ is not singular as well.
\medskip

\hspace{0em}From now on we consider vectors of Proposition~\ref{P:ss-q}.\\

7) $\,S=\lsp{\Dep(\pt_*)\pt_3^{q-1}\,,
             \Dem(\pt_*)\pt_3^{q-1}
                            }$. \\
We are to look at
\begin{eqnarray*}
\mx{\bf v}_1&=&\Dep(\pt_*)\pt_3^{q-1} \cdot z_+^{r}[\tht]\,,\\
\mx{\bf v}_2&=&\Dem(\pt_*)\pt_3^{q-1} \cdot z_+^{r}[\tht]
               -\Dep(\pt_*)\pt_3^{q-1}\cdot z_+^{r-1}z_-[\tht]\,.
\end{eqnarray*}
Let us notice that
\begin{align}
\label{eq:e0-Dp}
\/[e_0,\,\Dep(\pt_*)]=&\pt_1 (h_0-2) +\pt_2 f_1 +  \pt_3 f_{12}\,,\\
\/[e_0,\,\Dem(\pt_*)]=&-2 \pt_1  f_3 \,.\label{eq:e0-Dm}
\end{align}
Now it is easy to calculate that
\begin{align*}
e_0\mx{\bf v}_1=&((q-1)-\tht-2-(q-1)) \pt_1\pt_3^{q-1}\cdot z_+^{r}[\tht]\,,
\end{align*}
thus $\tht=-2$ for any $q\geq 1$, and we come to sigular vectors from ({\it 7}).

Similarly
\begin{align*}
e_0\mx{\bf v}_2=&
(-2r+\tht)\pt_1\pt_3^{q-2}\cdot z_+^{r-1}z_-[\tht]\,,\qquad\qquad\quad
\end{align*}
hence $\tht=2r$, and we come to sigular vectors from ({\it 8}).\\

8) $\,S=\lsp{ \dst^+_1\Dep(\pt_*)\,,
                 \dst^+_1\Dem(\pt_*),\,\dst^-_1\Dep(\pt_*)\,,
                   \dst^-_1\Dem(\pt_*)}$.\\
To simplify formulae we shall write just $\Dep, \, \Dem$
instead of $\Dep(\pt_*)$, $\Dem(\pt_*)$ whenever it does not lead to
misunderstanding.

For any $t\in T$ we have
\begin{align*}
\/e_0\,\dst^+_1\Dep\cdot t=&\s\s\Dep\cdot h_0 t -
\dst^+_1\pt_1\cdot (h_0-2)t\,,
\\
\/e_0\,\dst^+_1\Dem\cdot t=&\s\s
\Dem\cdot(h_0+2)t + 2 \dst^+_1\pt_1\cdot f_3t\,,
\\
\/e_0\,\dst^-_1\Dep\cdot t=&
-2\,\Dem\cdot t - \dst^-_1\pt_1\cdot (h_0-2)t
         -2\Dep\cdot f_3t\,,
\\
\/e_0\,\dst^-_1\Dem\cdot t=&
-2\,(\Dem-\dst^-_1\pt_1)\cdot f_3 t\,,
\end{align*}
Let us notice that the $s\ell(2)$-module $S$ is a direct sum of a
3-dimensional irreducible one and 1-dimensional one. The latter is
generated by $\dst^-_1\Dep(\pt_*)-\dst^+_1\Dem(\pt_*)$.
Therefore we have to consider the following vectors
\begin{eqnarray*}
\mx{\bf v}_1&=&\dst^+_1\Dep \cdot z_+^{r}[\tht]\,,\\
\mx{\bf v}_2&=&
(\dst^+_1\Dem+\dst^-_1\Dep)\cdot z_+^{r}[\tht]
               -2\dst^+_1\Dep\cdot z_+^{r-1}z_-[\tht]\,,\\
\mx{\bf v}_3&=& (\dst^+_1\Dem-\dst^-_1\Dep)\cdot z_+^{r}[\tht]\,,
\\
\mx{\bf v}_4&=& \dst^-_1\Dem \cdot z_+^{r}[\tht]
  -(\dst^+_1\Dem+\dst^-_1\Dep)\cdot  z_+^{r-1}z_-[\tht]
  +\dst^+_1\Dep\cdot z_+^{r-2}z_-^{2}[\tht]\,,
\end{eqnarray*}
and the linear combinations of $\mx{\bf v}_2,\,\mx{\bf v}_3$ as well.

Starting with $\mx{\bf v}_1$ we get
\begin{align*}
e_0\mx{\bf v}_1=&
-\tht\Dep\cdot z_+^{r}[\tht] + (\tht+2)\dst^+_1\pt_1\cdot z_+^{r}[\tht]\,.
\end{align*}
For the vector to be singular it should be $\tht=\tht+2=0$, which is
impossible.

Now from \eqref{eq:e0-Dp}, \eqref{eq:e0-Dm} we see that
\begin{align*}
e_0(\dst^+_1\Dem\cdot z_+^{r}[\tht])=&
             (-\tht+2)\Dem\cdot z_+^{r}[\tht]
            +2r\,\dst^+_1\pt_1\cdot z_+^{r-1}z_-[\tht]\,,\\
e_0(\dst^-_1\Dep\cdot z_+^{r}[\tht])=&
             -2\,\Dem\cdot z_+^{r}[\tht]
            +(\tht+2)\dst^-_1\pt_1\cdot z_+^{r}[\tht]
            -2r \Dep\cdot z_+^{r-1}z_-[\tht]\,,\\
e_0(-2\dst^+_1\Dep\cdot z_+^{r-1}z_-[\tht])=&
                 (2\tht-4)\Dep\cdot z_+^{r-1}z_-[\tht]
               -2\tht\,\dst^+_1\pt_1\cdot z_+^{r-1}z_-[\tht]\,.
\end{align*}
It follows that $e_0\,(\al\mx{\bf v}_2+\be\mx{\bf v}_3)=0$ if and only
if
\begin{eqnarray*}
\al(-\tht)+\be(2-\tht+2)&=&0\,,\\
(\al-\be)(\tht+2)&=&0\,,\\
\al(2\tht-4-2r)+\be\,2r&=&0\,,\\
\al(-2\tht+2r)+\be\,2r&=&0\,.
\end{eqnarray*}
\hspace{0em}From the second equation we see that either $\al-\be=0$ or $\tht=-2$.
It is easy to solve the system in each case. We get only one solution
with $r=1$, $\tht=2$, $\al=\be$ and that gives us the vector from
 ({\it 9}).

For the last vector $\mx{\bf v}_4 $ we calculate
\begin{align*}
e_0\,\,\dst^-_1\Dem\cdot z_+^{r}[\tht]\s =&
              -2r\,\Dem\cdot z_+^{r-1}z_-[\tht]
              +2r\,\dst^-_1\pt_1\cdot z_+^{r-1}z_-[\tht]\,,\\
e_0\,\dst^-_1\Dep\cdot z_+^{r-1}z_-[\tht]=&
              -2\,\Dem\cdot z_+^{r-1}z_-[\tht]
              +\tht\,\dst^-_1\pt_1\cdot z_+^{r-1}z_-[\tht]
              -2(r-1)\,\Dep\cdot z_+^{r-2}z_-^{2}[\tht]\,,\\
e_0\,\dst^+_1\Dem\cdot z_+^{r-1}z_-[\tht]=&
               -\tht\,\Dem\cdot z_+^{r-1}z_-[\tht]
               +2(r-1)\,\dst^+_1\pt_1\cdot z_+^{r-2}z_-^{2}[\tht]\,,\\
e_0\,\dst^+_1\Dep\cdot z_+^{r-2}z_-^{2}[\tht]=&
               \,(4-\tht)\Dep\cdot z_+^{r-2}z_-^{2}[\tht]
               -(2-\tht)\,\dst^+_1\pt_1\cdot z_+^{r-2}z_-^{2}[\tht]\,.
\end{align*}
and combine them to get  an explicite expression for
$e_0\,\mx{\bf v}_4=0$. We come to four linear equations on
$r,\,\tht$, that are inconsistent. Thus
no singular vector of this form exists. \\

9) $\,S=\lsp{ \Dem(\Dep(\pt_*)\pt_*)}$.\\
Here we are to compute
\begin{align*}
e_0\,\Dem(\Dep(\pt_*)\pt_*)\cdot z_+^{r}[\tht] =&
-2f_3\,\Dep \pt_1\cdot z_+^{r}[\tht]
- \sum \dst^-_i\, e_0\Dep\pt_i\cdot z_+^{r}[\tht]\\
=&-2\, \Dem\pt_1\cdot z_+^{r}[\tht]-2r\,\Dep \pt_1\cdot z_+^{r-1}z_-[\tht]\\
 +\left((\tht+2) \dst^-_1 \pt_1^2\right.+&
   \left.   \dst^-_2((3+\tht)\pt_1\pt_2-\pt_2\pt_1)+
  \dst^-_3((3+\tht)\pt_1\pt_3-\pt_3\pt_1)\right)
                                             \cdot z_+^{r}[\tht]\\
=&\s\tht\,\Dem\pt_1\cdot z_+^{r}[\tht]-2r\,\Dep \pt_1\cdot z_+^{r-1}z_-[\tht]\,.
\end{align*}
We come to conditions $r=0$, $\tht=0$ and then to the singular vector
listed in ({\it 10}).\\

10) $\,S=\lsp{\dst^-_1\dst^+_1 \Dep(\pt_*)}$.\\
Similarly we are to calculate
\begin{align*}
e_0\,\dst^-_1\dst^+_1 \Dep(\pt_*)\cdot z_+^{r}[\tht] =&
\left(-2f_3\,\dst^+_1 \Dep(\pt_*)
-\dst^-_1\,h_0\Dep(\pt_*)
+\dst^-_1\dst^+_1\,e_0\Dep(\pt_*)\right)\cdot z_+^{r}[\tht]\\
=&
\left((\tht-2)\dst^-_1 \Dep(\pt_*)
-2\dst^+_1 \Dem(\pt_*)
-(\tht+2)\dst^-_1\dst^+_1\right)\cdot z_+^{r}[\tht]\\
&-2r\,\dst^+_1 \Dep(\pt_*)\cdot z_+^{r-1}z_-[\tht]\, \neq 0\,.\\
\end{align*}
No singular vector exists.\\

11) $\,S=\lsp{ b\Dep,\, c\Dep,\, d\Dep}$.\\
Let us use the fact that $a\Dep=0$, thus $b\Dep+a\Dem=0$,
$c\Dep+b\Dem=0$, and $d\Dep+c\Dem=0$.
Therefore can consider vectors in the form
\begin{eqnarray*}
\mx{\bf v}_1&=&a\Dem\cdot z_+^{r}[\tht]\,,\\
\mx{\bf v}_2&=&b\Dem \cdot z_+^{r}[\tht]
               -2a\Dem\cdot z_+^{r-1}z_-[\tht]\,,\\
\mx{\bf v}_3&=&c\Dem \cdot z_+^{r}[\tht]
               -b\Dem\cdot z_+^{r-1}z_-[\tht]
                 +a\Dem\cdot z_+^{r-2}z_-^{2}[\tht]\,.
\end{eqnarray*}
Now for $\mx{\bf v}_1$, we use \eqref{eq:e0-a}, \eqref{eq:e0-Dm} and get
\begin{align*}
e_0\,\mx{\bf v}_1=&\left(
 [e_0,\,a]\Dep - a\,e_0\Dem
\right)\cdot z_+^{r}[\tht]\\
=&\left(\dst^+_2\dst^+_3(h_0-2)\Dem-a(-2f_3)\pt_1
\right)\cdot z_+^{r}[\tht]\\
=&
-\tht\dst^+_2\dst^+_3\Dem\cdot z_+^{r}[\tht]
+2r\,a\pt_1\cdot z_+^{r-1}z_-[\tht]\,.
\end{align*}
We conclude that $r=0$, $\tht=0$ and
this gives us the vector from ({\it 11}).

Similarly
\begin{align*}
e_0\,\mx{\bf v}_2=&\left(
 [e_0,\,b]\Dep - b\,e_0\Dem
\right)\cdot z_+^{r}[\tht]-2\left(
[e_0,\,a]\Dep - a\,e_0\Dem
\right)\cdot z_+^{r-1}z_-[\tht]
\\
=&\left(
-\tht(\dst^-_2\dst^+_3 + \dst^+_2\dst^-_3)\Dem
\right)\cdot z_+^{r}[\tht]\\
&+\left((-2r+2\tht-4)
\dst^+_2\dst^+_3\Dem + 2r\,b\pt_1
\right)\cdot z_+^{r-1}z_-[\tht]\\
&-4(r-1)a
\pt_1
\cdot z_+^{r-2}z_-^{2}[\tht]\,.
\end{align*}
We see that $e_0\,\mx{\bf v}_2\neq 0$ whatever $r,\,\tht$.

Again by the same argument we have
\begin{align*}
e_0\,\mx{\bf v}_3=&\left(
 [e_0,\,c]\Dep - c\,e_0\Dem
\right)\cdot z_+^{r}[\tht]
-\left(
 [e_0,\,b]\Dep - b\,e_0\Dem
\right)\cdot z_+^{r-1}z_-[\tht]\\
&+\left(
[e_0,\,a]\Dep - a\,e_0\Dem
\right)\cdot z_+^{r-2}z_-^{2}[\tht]\,.
\end{align*}
By $z_+^{r}[\tht]$ we get the coefficient
$\tht\,\dst^-_2\dst^-_3\Dem$, thus $\tht=0$.
The coefficient by $z_+^{r-1}z_-[\tht]$ gives
\begin{eqnarray*}
(-2r-(4-\tht-2))(\dst^-_2\dst^+_3 + \dst^+_2\dst^-_3)\Dem
 +2r\,c\pt_1&=&0\,,
\end{eqnarray*}
and from the coefficient by $z_+^{r-2}z_-^{2}[\tht]$ it comes
\begin{eqnarray*}
(2(r-1)+(6-\tht-2))\dst^+_2\dst^+_3\Dem
-2(r-1)\,b\pt_1&=&0\,.
\end{eqnarray*}
Clearly no singular vectors appear.\\

12) $\,S=\lsp{\dst^-_1\dst^+_1\Dem(\Dep(\pt_*)\pt_*)}$.\\
Here
\begin{eqnarray*}
\mx{\bf v}&=&\dst^-_1\dst^+_1\Dem(\Dep(\pt_*)\pt_*)\cdot z_+^{r}[\tht]\,,
\hspace{10em}
\end{eqnarray*}
It is enough to compute the coefficient by $z_+^{r}[\tht]$ in
$e_0\mx{\bf v}$ in order to realize that the result is non-zero. Namely
\begin{align*}
e_0\,\mx{\bf v}=&\left(
-2f_3\dst^+_1\Dem(\Dep(\pt_*)\pt_*)
-\dst^-_1\,h_0\Dem(\Dep(\pt_*)\pt_*)
+\dst^-_1\dst^+_1\,e_0\Dem(\Dep(\pt_*)\pt_*)
\right)
\cdot z_+^{r}[\tht]\\
=&\left(
(\tht-4)\dst^-_1\Dem(\Dep(\pt_*)\pt_*)
+ \tht \dst^-_1\dst^+_1\Dem\pt_1
\right)
\cdot z_+^{r}[\tht]+\ldots \neq 0\,.
\end{align*}
The vector is never singular.\\

13) $\,S=\lsp{\dst^-_1a\Dem\,,\dst^-_1b\Dem}$.\\
Now
\begin{eqnarray*}
\mx{\bf v}_1&=&\dst^-_1a\Dem\cdot z_+^{r}[\tht]\,,\\
\mx{\bf v}_2&=&\dst^-_1b\Dem \cdot z_+^{r}[\tht]
               -\dst^-_1a\Dem\cdot z_+^{r-1}z_-[\tht]\,.
\end{eqnarray*}
Immediately we get
\begin{align*}
e_0\,\mx{\bf v}_1=&\left(
(-2f_3)a\Dem -\dst^-_1[e_0,\,a]\Dem+\dst^-_1a\pt_1(-2f_3)
\right)\cdot z_+^{r}[\tht]\,,\\
=&(\tht\dst^-_1\dst^+_2\dst^+_3 -2\,b)
\Dem\cdot z_+^{r}[\tht]+ \ldots \neq 0\,.
\end{align*}
Also
\begin{align*}
e_0\,\mx{\bf v}_2=&\left(
(-2f_3)b\Dem -\dst^-_1[e_0,\,b]\Dem+\dst^-_1b\pt_1(-2f_3)
\right)\cdot z_+^{r}[\tht]+\ldots\\
=&(\tht\dst^-_1(\dst^-_2\dst^+_3 + \dst^+_2\dst^-_3)-4c)
\Dem\cdot z_+^{r}[\tht]+ \ldots \neq 0\,.
\end{align*}
There are no singular vectors here.\\

14) $\,S=\lsp{\hat{\De}(\dst^-_*)b\Dep,\,
\hat{\De}(\dst^-_*)c\Dep}$.\\
Again we can rewrite the basis using the relations
$b\Dep+a\Dem=0$, $c\Dep+b\Dem=0$. Thus we are to consider vectors
\begin{eqnarray*}
\mx{\bf v}_1&=&\hat{\De}(\dst^-_*)a\Dem\cdot z_+^{r}[\tht]\,,\\
\mx{\bf v}_2&=&\hat{\De}(\dst^-_*)b\Dem \cdot z_+^{r}[\tht]
               -\hat{\De}(\dst^-_*)a\Dem\cdot z_+^{r-1}z_-[\tht]\,.
\end{eqnarray*}
Here \eqref{eq:e0-D_}, \eqref{eq:e0-a} and \eqref{eq:e0-Dm} show
that
\begin{align*}
e_0\,\mx{\bf v}_1=&\left(
(-2\dst^-_2\dst^-_3-2\dhind 1f_3)a\Dem
-\hat{\De}(\dst^-_*)\dst^+_2\dst^+_3(h_0-2)\Dem
-2\hat{\De}(\dst^-_*)a\pt_1f_3
\right)\cdot z_+^{r}[\tht]\\
=&\left(
-2\dst^-_2\dst^-_3\,a\Dem -2\dhind 1 b\Dem
+\tht\hat{\De}(\dst^-_*)\dst^+_2\dst^+_3\Dem
\right)
\cdot z_+^{r}[\tht]+\ldots\\
=&\left(
2\dst^-_2\dst^-_3\,b\Dep +2\dhind 1 c\Dep
+\tht\hat{\De}(\dst^-_*)\dst^+_2\dst^+_3\Dem
\right)
\cdot z_+^{r}[\tht]+\ldots\,.
\end{align*}
It is clear that when we put the terms on the right hand side in
order, we get only one term not containing
$\dhind 1, \dhind 2, \dhind 3$, namely
$2\dst^-_2\dst^-_3\dst^-_1\dst^+_2\dst^+_3\dst^+_1\pt_1\cdot z_+^{r}[\tht]$,
hence $e_0\,\mx{\bf v}_1 \neq 0$.

Calculating with $e_0\,\mx{\bf v}_2$ we shall check only terms with
$z_+^{r}[\tht]$. Then
\begin{align*}
e_0\,\mx{\bf v}_2=&\left(
(-2\dst^-_2\dst^-_3-2\dhind 1f_3)b\Dem
-\hat{\De}(\dst^-_*)(\dst^-_2\dst^+_3 + \dst^+_2\dst^-_3)(h_0-2)\Dem
\right)\cdot z_+^{r}[\tht]+\ldots\\
=&\left(
2\dst^-_2\dst^-_3\,c\Dep+4\dhind 1 d\Dep
+\tht\hat{\De}(\dst^-_*)(\dst^-_2\dst^+_3 + \dst^+_2\dst^-_3)
\Dem
\right)
\cdot z_+^{r}[\tht]+\ldots\,.
\end{align*}
Notice that
$\dst^-_2\dst^-_3\,c=\dst^-_2\dst^-_3\dst^+_2\dst^-_3\dst^-_1
=-\dhind 1\dst^-_2\dst^-_3\dst^-_1=-\dhind 1\,d$. Therefore
\begin{align*}
e_0\,\mx{\bf v}_2=&\left(
3\dhind 1 d\Dep
+\tht\hat{\De}(\dst^-_*)(\dst^-_2\dst^+_3 + \dst^+_2\dst^-_3)
\Dem
\right)
\cdot z_+^{r}[\tht]+\ldots\, \neq 0
\end{align*}
We get no singular vectors in this case and we have come an end in
the list of cases. Thus we have found all the singular vectors listed and
nothing else,
this ends the proof of Theorem~\ref{T:sing-v}.
\end{proof}

%

\vspace{6ex}

\textbf{Authors' addresses:}
\begin{list}{}{}

\item  Department of Mathematics, MIT,
Cambridge MA 02139,
USA,\\
email:~~kac@math.mit.edu

\vspace{1ex}
\item   Inst. for Matemat. Fag, NTNU, Gl\o shaugen,
7491 Trondheim,
NORWAY, \\
email:~~rudakov@math.ntnu.no
\end{list}
\end{document}